\documentclass[final,onefignum,onetabnum]{siamart220329}

\usepackage{graphicx}
\usepackage{dcolumn}
\usepackage{bm}

\usepackage[utf8]{inputenc}
\usepackage{float}
\usepackage{mathtools}
\usepackage{tikz}
\usepackage{fullpage}
\DeclareMathOperator{\Tr}{Tr}
\usepackage{amsfonts}

\usepackage{capt-of}

\usepackage{comment}
\usepackage[normalem]{ulem}

\newcommand{\etal}{\emph{et al.}}
\begin{document}

\title{Analysis of mean-field approximation for Deffuant opinion dynamics on networks}
\author{
Alina Dubovskaya\thanks{Department of Mathematics and Statistics, University of Limerick, Ireland (\email{alina.dubovskaya@ul.ie}, \email{kevin.burke@ul.ie}, \email{james.gleeson@ul.ie}, \email{doireann.okiely@ul.ie}, \url{http://www.doireannokiely.com}).}
\and
Susan C. Fennell\footnotemark[1]
\and
Kevin Burke\footnotemark[1]
\and 
James P. Gleeson\footnotemark[1]
\and
Doireann O'Kiely\footnotemark[1]
}

\maketitle 

\begin{abstract}
Mean-field equations have been developed recently to approximate the dynamics of the Deffuant model of opinion formation. These equations can describe both fully-mixed populations and the case where individuals interact only along edges of a network. In each case, interactions only occur between individuals whose opinions differ by less than a given parameter, called the confidence bound. The size of the confidence bound parameter is known to strongly affect both the dynamics and the number and location of opinion clusters. In this work we carry out a  mathematical analysis of the mean-field equations to investigate the role of the confidence bound and boundaries on these important observables of the model. 
We consider the limit in which the confidence bound interval is small, and identify the key mechanisms driving opinion evolution.  We show that linear stability analysis can predict the number and location of opinion clusters.  Comparison with numerical simulations of the model illustrates that the early-time dynamics and the final cluster locations can be accurately approximated for networks composed of  two degree classes, as well as for the case of a fully-mixed population.
\end{abstract}

\begin{keywords}
opinion dynamics, mean-field approximation, linear stability, asymptotic analysis, cluster formation, social networks
\end{keywords}

\begin{MSCcodes}
91D30, 35R09, 35Q91
\end{MSCcodes}

\section{Introduction}
\subsection{Opinion dynamics}
Social interaction is a driving force in shaping people’s opinions, beliefs, and behaviours. The mechanisms underpinning these opinion dynamics are of key interest as policy-makers seek to encourage individuals to adopt more environmentally-friendly practices~\cite{Ireland2020,Ireland2021}, and to understand social phenomena such as vaccine hesitancy~\cite{dube2013}. Models of opinion dynamics seek to explain macro-level phenomena, such as global consensus or local opinion clustering, by specifying simple rules for how people interact at a micro level~\cite{Flache2017}. 
Some of the earliest models for opinion dynamics demonstrated that groups form a consensus when individuals seek to become more similar (i.e., move their opinion closer) to those they interact with~\cite{DeGroot1974, French1956,Harary1959}.  In the real world, opinions do not always reach a consensus, as evidenced by political division in numerous countries in the last decade~\cite{Ford2017,Abramowitz2018,Layton2021}. 
Theories from psychology and data from social media platforms both suggest that individuals interact preferentially with people whose opinions are similar to their own~\cite{Axelrod1997,Bakshy2015,Conover2011,Nikolov2015}. From a mathematical modelling point of view, the first mechanism shown to produce polarization on networks was bounded confidence, in models proposed by Krause~\cite{Krause2000} and Deffuant et al.~\cite{Deffuant2000}.  Under
bounded confidence, {an individual's opinion is a scalar value in the interval $[0,1]$ and it }is influenced only by {interactions with} those people who have
opinions close enough to, i.e., within a distance $\epsilon$ of, their own. Depending on the value of $\epsilon$, these models can produce consensus, as in the classical averaging models, or polarization.  Models with repulsive influence were introduced to explain how opinion differences could increase over time, leading to the {population} splitting into two factions with opposing views~\cite{Macy2003,Jager2005,Salzarulo2006,Flache2017}. 

The effect of network structure on dynamical processes such as epidemic spreading and information diffusion is an important area of research in network science~\cite{Newman2010,Porter2016}.  Many opinion dynamics models were originally
studied on fully connected graphs {(the ``fully-mixed'' population case) and on} lattices. These networks are not representative of real world networks which exhibit features such as low average path length (the “small world effect”)~\cite{Albert1999,Milgram1967,Watts1998}, community structure~\cite{Ferrara2012,Girvan2002,Newman2006,Shai2021} and heterogeneous degree distributions~\cite{Barabasi1999,Clauset2009}. In addition to agent-based simulations, mean-field approximations are useful for studying dynamics that take place on the nodes of a (large) network. By making assumptions about the network structure and dynamical correlations~\cite{Porter2016}, a set of equations for the time-dependent proportion of agents with a given opinion can be derived. The number of such equations is typically much smaller than the system size and so numerical integration of those equations is more efficient than Monte Carlo simulations of the entire system. 

In this paper we carry out a detailed mathematical analysis of a mean-field approximation of {bounded-confidence} opinion dynamics on a network. We address both {fully connected} networks and networks with different degree classes, and 
we use mathematical analysis to explain how opinions evolve on the network and how opinion clusters form.

\subsection{Deffuant model}\label{sec:deffuant_mf_analysis_intro}
\begin{figure}[ht]
\centering
\begin{tikzpicture}
\node at (0,0) {\includegraphics[width=13cm]{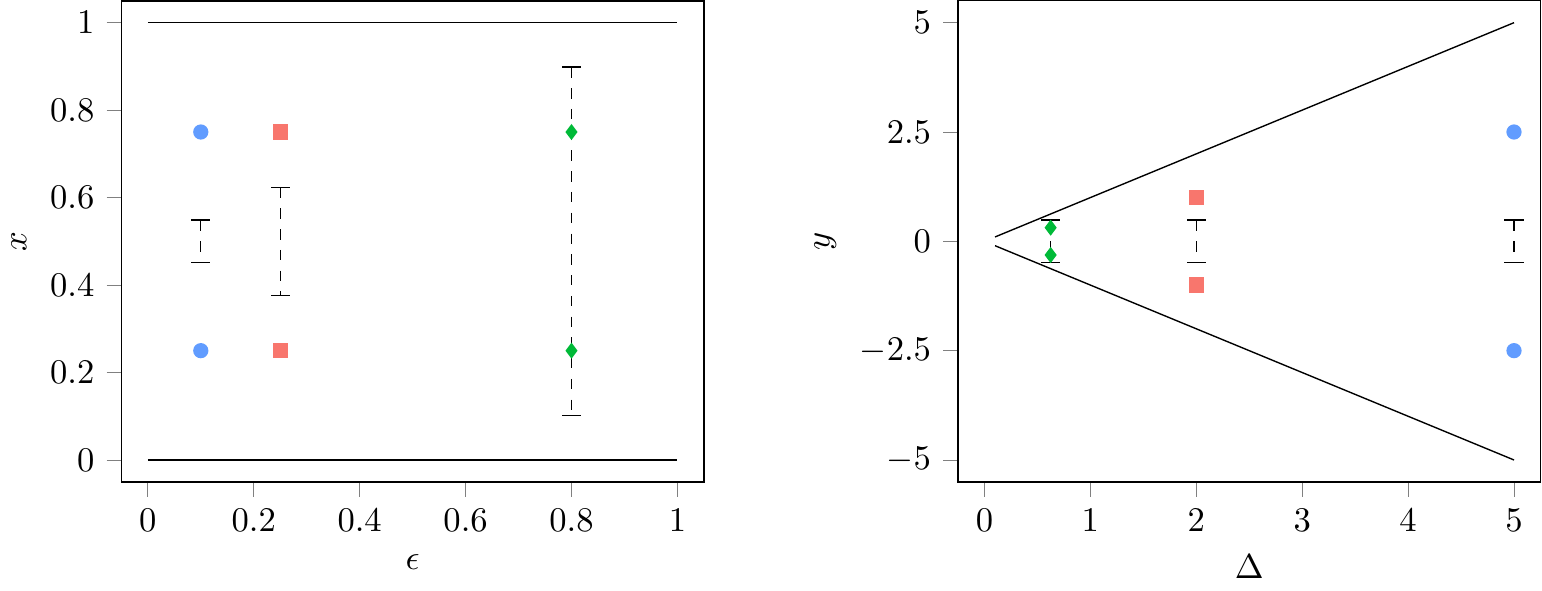}};
\node at (-6.5,2.3) {(a)};
\node at (0.3,2.3) {(b)};
\end{tikzpicture}
\caption{\label{fig:schematic_opinion_space}
Deffuant dynamics: (a) In $(x,\epsilon)$ space, individuals have opinions $\{x_i\} \in [0,1]$ and two connected individuals can interact if their opinion difference $|x_i-x_j| < \epsilon$.  In this illustration, two nodes with opinion difference {of} $0.5$ cannot interact if the confidence bound $\epsilon<0.5$, but can interact when the confidence bound $\epsilon = 0.8$.  (b) In $(y, \Delta)$ space the confidence bound is fixed while the opinion space widens as $\Delta = 1/2\epsilon$ increases. }
\end{figure}

We focus on the Deffuant model, which can simulate both consensus and opinion clustering.  In this model, the mechanism underlying opinion evolution is as follows. Individuals $i \in \{1,\cdots,N\}$ have {time-dependent} opinions $x_i \in [0,1]${, with the initial distribution of opinions typically being uniform on $[0,1]$}.  These individuals are nodes on a network, and a node $i$ may be connected to a node $j$ {by} an edge $(i,j)$.  The number of { edges connected to  node $i$ is denoted $k_i$ and is referred to as the degree of node $i$}. {At each {discrete} time step $n$, a pair of connected nodes $i$ and $j$ is chosen at random. If the difference $|x_i-x_j|$ between the nodes' opinions is less than a confidence bound $\epsilon$, the nodes interact and update their respective opinions. {In this paper, we focus on the scenario where an interacting pair each arrive at a new opinion which is half-way between the pair's previous opinions, i.e.}
\begin{equation}\label{eq:update}
        x_i(n+1) = x_j(n+1)= \dfrac{ x_j(n)+x_i(n)}{2}, \quad \text{if} \; | x_i(n)-x_j(n)| <\epsilon. 
\end{equation}
{This corresponds to a weighting $\mu = 1/2$ in the language of~\cite{Deffuant2000}. No }updates in opinions occur if $|x_i(n)-x_j(n)| \geq\epsilon$. 
An alternative formulation of the model was suggested in~\cite{Ben-Naim2003} in term of opinions $y$ {defined on an opinion space where the confidence bound is fixed at 1 but the space expands as the confidence bound decreases}. The transformation from $(x,\epsilon)$ space to the new space is {obtained} via
\begin{equation} \label{eq:transformation}
y_i = \dfrac{2x_i-1}{2\epsilon}, \quad \Delta = \dfrac{1}{2\epsilon}, \quad \text{where} \; y_i \in [-\Delta, \Delta],
\end{equation}
and the update rule~\eqref{eq:update} in $(y,\Delta)$ space is
\begin{equation}
    \begin{array}{cc}
        y_i(n+1) = y_j(n+1)= \dfrac{y_j(n)+y_i(n)}{2}, \quad \text{if} \; | y_i(n)-y_j(n)| <1. \\
    \end{array}
\end{equation}
The original $(x,\epsilon)$ parametrization of the model is natural and can be easily extended to other scenarios, for instance when each individual $i$ has their own confidence bound $\epsilon_i$.  However, the alternative $(y,\Delta)$ formulation can be helpful for interpreting model results. Therefore, we frame our analysis in $(x, \epsilon)$ space, but often utilize $(y, \Delta)$ space for illustrative purposes. Figure~\ref{fig:schematic_opinion_space} shows {a comparison} between the two spaces.}

\subsection{Mean-field approximation of the Deffuant model}
While the Deffuant model has mostly been studied via Monte Carlo simulations of the dynamics, mean-field approximations have been developed and analysed by Ben-Naim \etal~\cite{Ben-Naim2003} and by a subset of the current authors~\cite{Fennell2021}. Instead of keeping track of how each individual changes their opinion over time, the mean-field approach is to consider the limit $N \to \infty$ and derive rules for the density $P(x, t)$ of opinions $x\in[0,1]$ and how this evolves in time $t$.  In~\cite{Ben-Naim2003}, a fully-mixed population was assumed; this means that each agent may interact with any other agent whose opinion is close enough to their own. 

A more general mean-field approximation was developed in~\cite{Fennell2021} to describe how Deffuant dynamics are affected by the network structure of large networks of configuration-model type~{\cite{Newman2010}}. The {population-level density $P(x, t)$  of~\cite{Ben-Naim2003} } was extended to multiple opinion densities $P_k(x, t)$, one for each distinct degree class $k$ (i.e., group of nodes with the same degree $k$) in the network. {Numerical solutions were presented} for the special case of two classes of nodes with two different degrees of connectivity:  one group {of nodes} with {degrees equal to 100} and a second {larger} group whose {degrees are much smaller}.  For a confidence bound $\epsilon = 0.3$, most of the ``mass'' of the population formed a consensus near the centre of the opinion space, but some mass is ``left behind'' near the boundaries, with the proportion of the population in these boundary regions increasing in less connected networks. 
{ A further generalization considered other possible partitions} of the network nodes into discrete classes, with the probability for two nodes being connected on the network depending only on their class labels.  This allows other useful partitions to be considered, for example networks with community structure{~\cite{Newman2010}}.

\subsection{Paper outline}
In this paper, we use a range of mathematical techniques to obtain analytical results both for fully connected networks and for {networks with} multiple degree classes.  In the case of multiple degree classes, we present results for {an example that was the main focus of}~\cite{Fennell2021}: 10\% of the population have degree 100 and the other 90\% (i.e.\ the majority) of the population have degree  $k_{maj}<100$.  
We start by introducing the mean-field approximation of the Deffuant model with multiple degree classes in  Sec.~\ref{sec:meanfield}; the well-mixed system of~\cite{Ben-Naim2003} is a special case of this.  We carry out a preliminary analysis, demonstrating that the solution behaves differently in different regions in the opinion space and explaining why this happens.  The major mathematical analysis of the paper then has two parts, an analysis of the regime $\epsilon \ll 1$ in Sec.~\ref{sec:omf_numerical_approx} and a linear stability analysis in Sec.~\ref{sec:linearstability}.  

The regime $\epsilon \ll 1$ corresponds to a small confidence bound, where interactions {occur only} over a limited opinion range and the {distribution of opinions} {therefore evolves} slowly.  If the opinion density is initially uniform, then the density of opinions in the centre of opinion space are fixed at leading order, and opinion evolution is driven by an imbalance in interactions near the boundaries.  Near the boundaries, {we show that }opinions undergo a canonical evolution that is simply rescaled for different $\epsilon$.  The observation that a uniform opinion field does not evolve away from the influence of boundaries motivates our linear stability analysis of the mean-field equations {in Sec.~\ref{sec:linearstability}}.  This linear stability analysis enables us to estimate the size and location of opinion clusters observed in numerical solutions of the full mean-field problem. For both methods of analysis, we apply our results to the case {of a large network} with two degree classes. We conclude with a discussion in Sec.~\ref{sec:mf_analysis_conclusions}.

\section{Class-based mean-field model formulation}
  Following the formulation of~\cite{Fennell2021}, nodes are categorized according to their degree into \emph{degree classes}.  If two nodes interact --- which requires them to be connected by a network edge as well as to have opinions that differ by less than $\epsilon$ --- their opinions are updated according to an averaging rule. The probability density function for the opinions of individuals in each degree class $k$ is denoted $P_k(x,t)$, and the $\{P_k\}$ are governed by a system of coupled integro-differential equations
\begin{equation}\label{eq:chapter2}
    \frac{\partial}{\partial t} P_k(x,t) = \sum_l 
    C_{kl}
    \left[ 
    \int_{|z|<\epsilon} P_k(x+z/2,t) P_l\left(x-z/2, t \right)\ \mathrm{d} z
    - 
    \int_{|z|<\epsilon} P_k(x,t) P_l(x+z,t)\ \mathrm{d} z
    \right],
\end{equation}
where the coefficient $C_{kl} = q_l \pi_{kl}/\gamma$, $q_l$ is the fraction of nodes in the degree class $l$, $\pi_{kl} = k l / N \sum (kq_k)$ is the probability that an edge exists between two nodes drawn at random from classes $k$ and $l$, and $\gamma = 2|E|/N^2$ is the graph density with $|E| = \sum_k \sum_l q_k q_L \pi_{kl}$ the total number of edges in the network.  The first integral on the right-hand side quantifies the increase in the number of individuals with an opinion $x$ that occurs when two individuals with differing opinions $x-z/2$ and $x+z/2$ interact, provided the distance $|z|$ between their opinions is less than the confidence bound $\epsilon$.  The second integral quantifies the decrease in the number of individuals with an opinion $x$ that occurs when an individual with opinion $x$ interacts with an individual with a different opinion $x+z \in (x-\epsilon, x+\epsilon)$.  Note that we have measured time as outlined in~\cite{Fennell2021}, so that in one unit of time every node will update their opinion once on average.  We note also that some authors measure time in such a way that in one unit of time every node updates their opinion twice on average~\cite{pineda2009, Gomez2012}; this corresponds to rescaling $t$ by a factor 2 in~\eqref{eq:chapter2} but otherwise has no effect on the dynamics of the system.

We observe that at a given {opinion value} $x^*$, the contributions from the two integrals can cancel to give $\partial P/\partial t = 0$ if $P \equiv 1$ in the interval $[x^*-\epsilon,x^*+\epsilon]$.  If $P$ is not identically 1 in this interval, or if the interval is restricted by the size of the opinion space, then the integrals do not balance and $\partial P/\partial t \neq 0$.  
This suggests that $P \equiv 1$ would be a steady-state solution on a boundary-free domain, and we use this as the basis of our linear-stability analysis in \S\ref{sec:linearstability}.  We also suggest that, when $\epsilon \ll 1$, there may be a large section of the domain, at distances greater than $\epsilon$ from the boundaries, where $P$ changes very slowly.  Networks with small confidence bounds may therefore be more amenable to mathematical investigation than when $\epsilon = O(1)$.  We carry out this analysis in \S\ref{sec:omf_numerical_approx}, but we first use numerical and mathematical analysis to identify some key features of the system.

\section{Numerical results}\label{sec:meanfield}
{In this section, we discuss valuable insights obtained from numerical results which we will use later in mathematical analysis. We present the results for the fully mixed mean-field model, however, our finding are equally applied to the class-based model. }

\subsection{Late-time dynamics}
We begin by discussing the $t\to\infty$ solution of the fully-mixed mean-field model of Ben-Naim \etal~\cite{Ben-Naim2003}
\begin{equation}
\label{eqn:omf}
\begin{split}
\frac{\partial}{\partial t}P(x,t) = \underbrace{\int_{|z|<\epsilon}P(x+z/2,t)P\left(x-z/2,t\right)\,dz}_{I_1} -\underbrace{\int_{|z|<\epsilon}P(x,t)P(x+z,t)\,dz}_{I_2},
\end{split}
\end{equation}
which is a special case of~\eqref{eq:chapter2}.
To solve this equation numerically we adopt the method of Lorenz and co-authors~\cite{lorenz2007thesis,lorenz2007continuous} and rewrite it as a Markov chain with the opinion space $x \in [0,1]$ divided into $200/\epsilon$ opinion classes.  The integrals are then evaluated through transition probabilities for mass transitioning into a given class and out of a given class. Initially, the opinions are distributed uniformly, i.e. $P(x,0) = 1$.

\begin{figure}[ht]
\centering
\begin{tikzpicture}
\node at (0,0) {\includegraphics[trim = 0mm 30mm 0mm 0mm, clip, width=8cm]{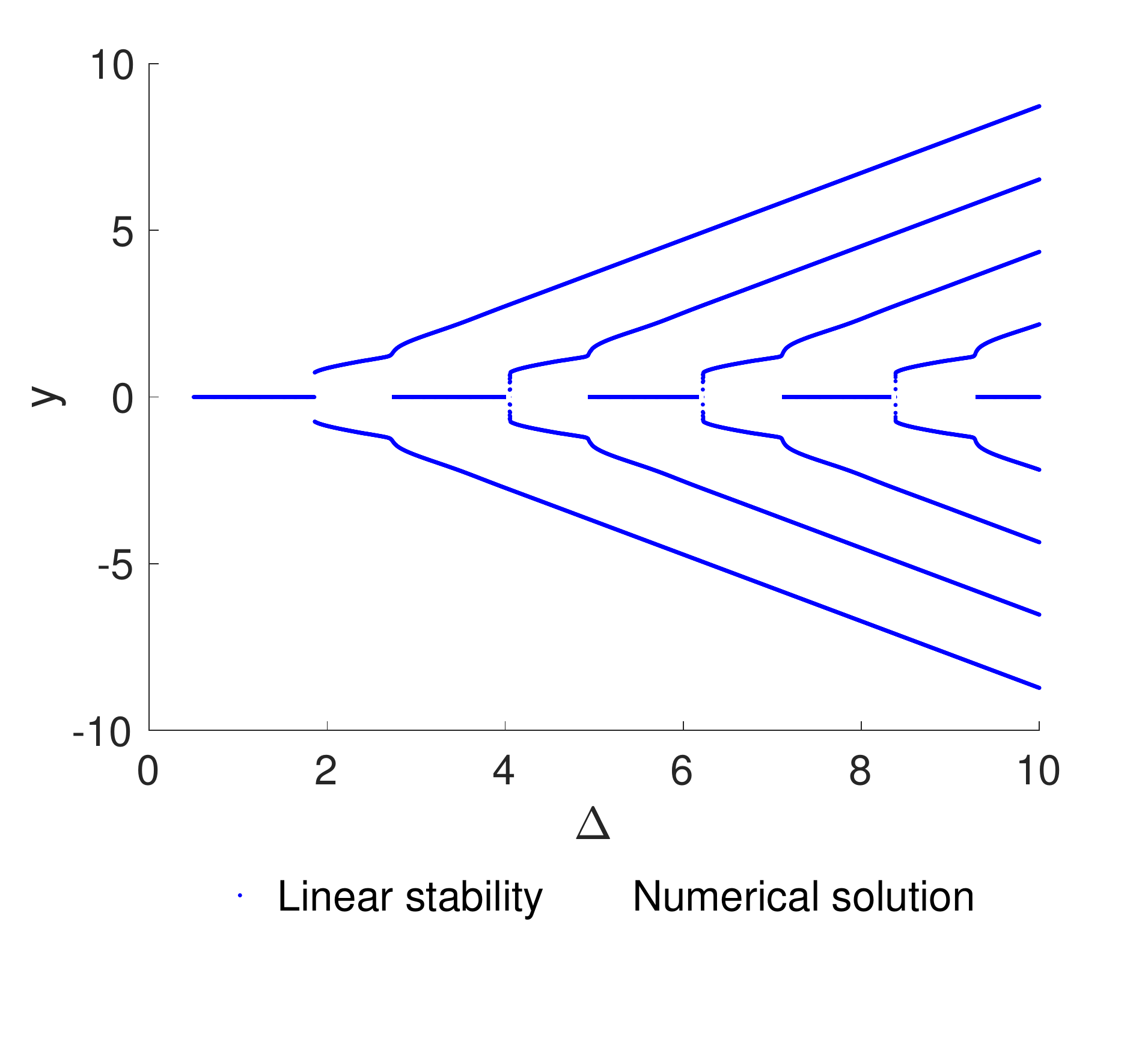}};
\node at (8.0,0) {\includegraphics[trim = 0mm 30mm 0mm 0mm, clip, width=8cm]{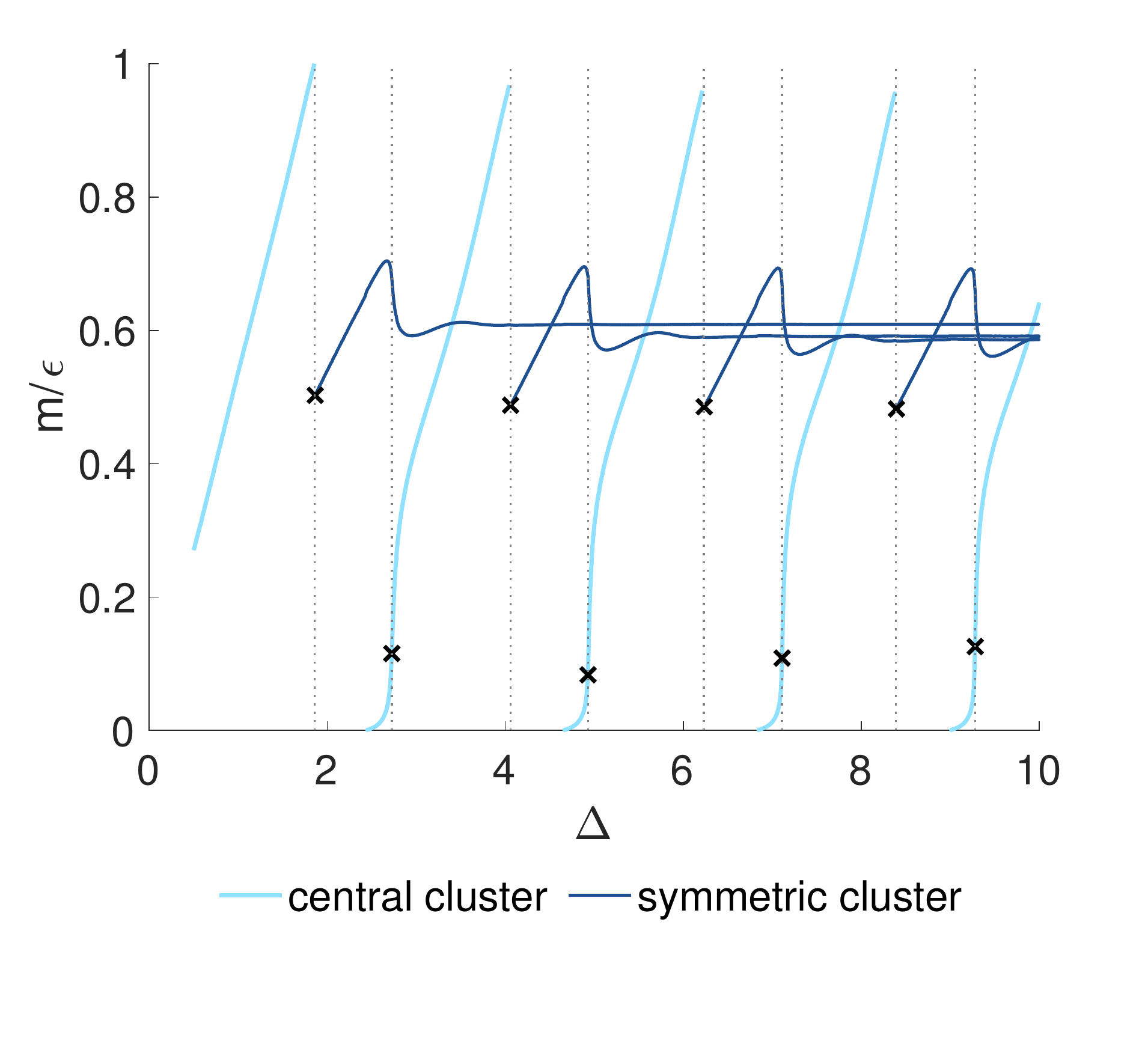}};
\node at (-3.8,3) {(a)};
\node at (4.3,3) {(b)};
\end{tikzpicture}
\caption{\label{fig:bif_diag_num} (a) Location of peaks from numerical solution of the {mean-field} equation~\eqref{eqn:omf} in  $(y,\Delta)$ space. This is a reproduction of Fig.~2 in~\cite{Ben-Naim2003}, but with focus on major clusters only and a modified identification method for bifurcation points. (b) Mass  of  the opinion  clusters from numerical simulations. The light blue lines  are  the central clusters and the dark blue lines correspond to symmetric clusters. The mass of each cluster is proportional to $\epsilon$ and therefore the clusters get very small as $\epsilon$ gets small ($\Delta$ gets large); for this reason, we plot the mass divided by $\epsilon$. The dotted lines mark values of $\Delta$ corresponding to bifurcation points; these are the values of $\Delta$ where the growth rate of a cluster reaches its maximum (marked with crosses).  These locations also correspond to  sharp changes in masses of neighbouring symmetric clusters.}
\end{figure}

The Deffuant dynamics eventually lead to formation of opinion clusters which, in the long-time limit solution of $P(x,t)$, are represented by delta functions located at positions separated by the distances greater than $\epsilon$. The number of those clusters is  defined by the value of confidence bound. It is therefore of interest to {examine} the number and positions of the clusters for a range of values of the parameter $\Delta$.
As Fig.~\ref{fig:bif_diag_num}a shows, {in $(y,\Delta)$ space} the number of opinion clusters increases with increasing $\Delta$ in a pattern that repeats every time $\Delta$ increases by approximately 2 units. There are two types of bifurcations in the diagram: in the first type, two symmetric clusters are created with simultaneous annihilation of a central cluster; in the second type, the central cluster reappears again. The symmetric clusters appear with a large mass as Fig.~\ref{fig:bif_diag_num}b shows. However, the central cluster initially emerges with a small mass which then rapidly grows as $\Delta$ increases.

The cyclic appearance, growth and splitting of the central cluster was previously outlined by~\cite{Ben-Naim2003}.  As $\Delta$ is increased, a central cluster appears with a very small mass; the mass of the cluster initially increases slowly as $\Delta$ increases and then increases rapidly at a particular value of $\Delta$ (see e.g. the light blue lines in Fig.~\ref{fig:bif_diag_num}b).  As $\Delta$ is increased further, the central cluster splits and then the cycle repeats.  In~\cite{Ben-Naim2003}, the bifurcation points for the central clusters were defined based on the first value of $\Delta$ at which a new central cluster of very small mass was detected in the large-time limit of a numerical solution of the mean-field equation, and the cluster location is dependent on the threshold mass for detection. In this study, we instead defined the points of bifurcation of central clusters to be the values of $\Delta$ where the mass of the central cluster increases rapidly (marked in the Fig.~\ref{fig:bif_diag_num}b with crosses).  In particular, we identify the values of $\Delta$ where the gradient $\mathrm{d}(m/\epsilon)/\mathrm{d}\Delta$ is maximized. To justify our choice, we note that  bifurcations of central clusters {defined in this way} align with the sharp shifts in the position of the nearest symmetric clusters in Fig.~\ref{fig:bif_diag_num}a, as well as with a rapid decrease in the {mass of the symmetric clusters  (see Fig.~\ref{fig:bif_diag_num}b).}

\subsection{Early-time dynamics}
\begin{figure}[ht]
\centering
\begin{tikzpicture}
\node at (0,0) {\includegraphics[trim = 0mm 0mm 0mm 0mm, clip, width=8cm]{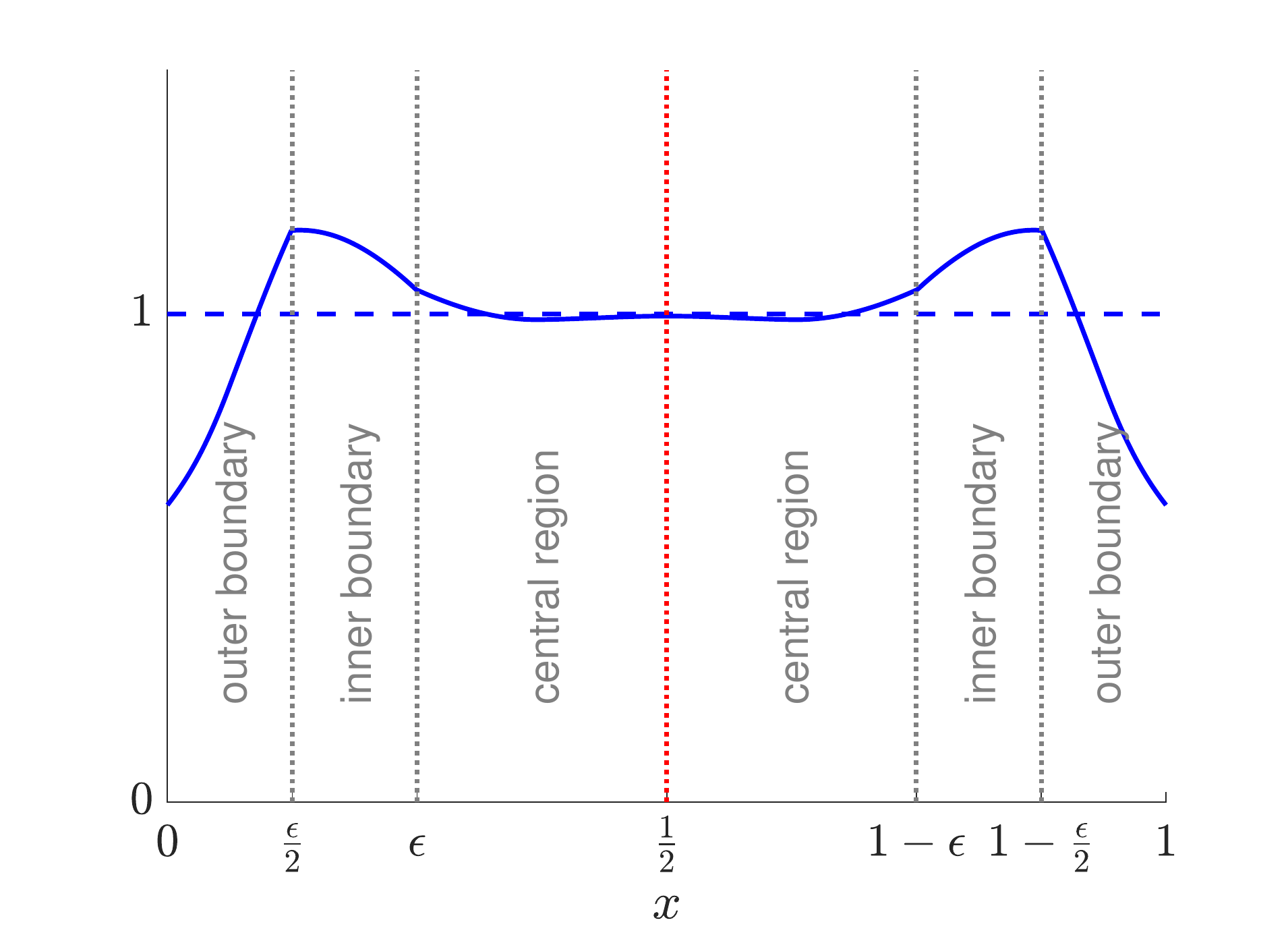}};
\node at (8.0,0) {\includegraphics[trim = 0mm 0mm 0mm 0mm, clip, width=8cm]{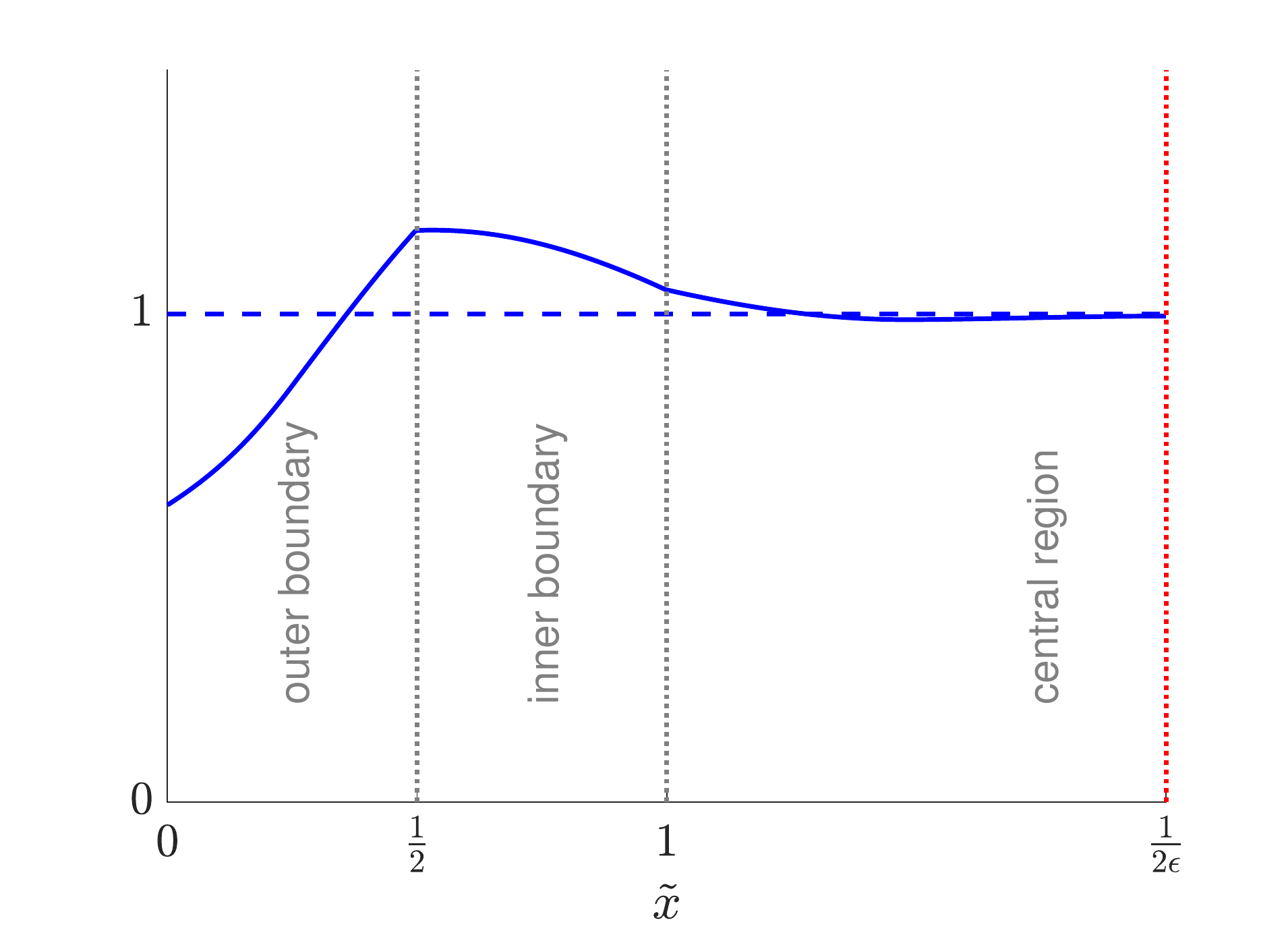}};
\node at (-3.8,3) {(a)};
\node at (4.3,3) {(b)};
\end{tikzpicture}
\caption{\label{fig:regions_new} Typical early-time solution of the mean-field equation~\eqref{eqn:omf} (a) in $(x,\epsilon)$ space; (b) in $\tilde{x}$ space. Here $\epsilon = 0.25$ and $t=2$. Five distinct $x$ regions are mapped into three in $\tilde{x}$ space. The initial distribution $P=1$ is included (dashed line) for reference.}
\end{figure}

We now consider the early-time dynamics of the mean-field equation~\eqref{eqn:omf} and discuss how the opinion clusters emerge from the uniform initial distribution $P(x,0)=1$. An example solution is illustrated in Fig.~\ref{fig:regions_new}a for $\epsilon = 0.25$ at early time $t=2$. Five distinct {$x$} regions are clearly apparent. We characterize these regions in terms of the probability mass
describing the density of people in different parts of the opinion space $[0,1]$ and their motion within this space.  Nearest to the two boundaries $x = 0,1$ are zones $[0,\epsilon/2)$ and $(1-\epsilon/2,1]$ where mass flows away from the boundaries and towards the centre of the opinion space.  Initially, this mass accumulates in two inner zones $[\epsilon/2, \epsilon)$ and $(1-\epsilon,1-\epsilon/2]$.  The distribution in the central region $[\epsilon,1-\epsilon]$ is almost unchanged at early times. We can rationalize the appearance of these regions by considering  the limits of each integral in~\eqref{eqn:omf}.

\subsection{Qualitative rationalization} \label{sec:omf_2}
We denote the two integrals in~\eqref{eqn:omf} by $I_1$ and $I_2$ respectively.  The domain of integration in $I_1$ is $|z|<\epsilon$, but $P(\xi,t)$ is only defined when the dummy variable $\xi \in [0,1]$, so when $x < \epsilon/2$ the integral is actually taken over $|z|<2x$. Analogous restrictions apply {when} $x > 1-\epsilon/2$ for $I_1$ and {when} $x<\epsilon$ and $x>1-\epsilon$ for $I_2$, giving rise to the five regions noted above.  The limits for each region are summarized in Table~\ref{table:integral_limits} and can help us to understand the distribution shape in each region, at least at early times.

\begin{table}[ht]
\setlength\tabcolsep{3.5pt}
\begin{tabular}{cccccc}
\hline \hline
 && \multicolumn{2}{c}{$I_1$}&\multicolumn{2}{c}{$I_2$} \\ \hline
\multicolumn{2}{c}{Region} &lower limit $z_{l_1}$ & upper limit $z_{u_1}$ & lower limit $z_{l_2}$ & upper limit $z_{u_2}$ \\ \hline
Outer boundary &$\left[0,\frac{\epsilon}{2}\right)$  & $-2x$ & $2x$ & $-x$ & $\epsilon$ \\
Inner boundary &$\left[\frac{\epsilon}{2},\epsilon\right)$ & $-\epsilon$ & $\epsilon$ & $-x$ & $\epsilon$  \\
Central&$\left[\epsilon, 1-\epsilon\right]$ & $-\epsilon$ & $\epsilon$ & $-\epsilon$ & $\epsilon$  \\
Inner boundary &\hspace{-7pt}$\left(1-\epsilon,1-\frac{\epsilon}{2}\right]$ & $-\epsilon$ & $\epsilon$  & $-\epsilon$ & $1-x$ \\
Outer boundary &$\left(1-\frac{\epsilon}{2},1\right]$ & $-2(1-x)$ & $2(1-x)$ & $-\epsilon$ & $1-x$ \\
\hline \hline
\end{tabular}
\vspace{2mm}
\caption{Integral limits in each region for Equations~\eqref{eq:chapter2} and \eqref{eqn:omf}. }
\label{table:integral_limits}
\setlength\tabcolsep{6pt}
\end{table}

Consider the outer boundary region $[0,\epsilon/2)$. As $x\rightarrow 0 $ the domain of integration in $I_1$ tends to 0 and so $I_1 \rightarrow 0$. The domain of integration in $I_2$ also decreases as $x\rightarrow 0$, however the integral is always over  a domain of at least length $\epsilon$. This means that, at least for early times, as $x\rightarrow0$ the negative contribution to $\partial P/\partial t$ outweighs the positive contribution, and so the probability density decreases for small $x$ as observed in Fig.~\ref{fig:regions_new}a. The mass leaving the zone $[0,\epsilon/2)$ must flow towards larger $x$, into the zone $(\epsilon/2,\epsilon)$. The mass at a position $x \in (\epsilon,1-\epsilon)$ only changes if $P\neq 1$ in {the interval} $(x-\epsilon,x+\epsilon)$, so the distribution changes more slowly near the centre of the domain.  Overall, there is a flux of mass to the right from the zone $(0,\epsilon/2]$ which cannot travel the whole way to the centre of the domain at early time, resulting {in a} local peak near $x = \epsilon/2$ and{, analogously, another} near $x = 1-\epsilon/2$.

\subsection{Symmetry}
Figure~\ref{fig:regions_new}a also suggests that the distribution is symmetric about $x=1/2$.  In fact, if we start with {any} symmetric initial distribution $P(x,0) = P(1-x,0)$ then the distribution will be symmetric for all time. We can verify this must be the case by comparing ${\partial P}/{\partial t}$ at two points $x$ and $1-x$.  In particular, if $P$ is symmetric about $x=1/2$ then the right-hand-side of~\eqref{eqn:omf} is the same at $x$ and $1-x$.  This means that if the distribution is symmetric at time $t$ then it must be symmetric at time $t+dt$ for an infinitesimal time step $dt$, and by extension if the distribution is symmetric at $t=0$ then it must be symmetric for all time.

\section{Small confidence bound analysis} \label{sec:omf_numerical_approx}
We now consider the case $\epsilon \ll 1$, in which the confidence bound interval is small and individuals only interact with people whose opinions are very similar to their own.  We anticipate that opinions will evolve more slowly in this regime where the number of interactions per unit time is lower.  Examining the degree-based governing equation~\eqref{eq:chapter2} and fully-mixed governing equation~\eqref{eqn:omf}, we see that in each case the ranges of the integrals quantifying movement to and away from an opinion $x$ shrink as $\epsilon \to 0$.  
Based on the occurrence of regions near the boundaries with altered integral limits in Table~\ref{table:integral_limits}, we anticipate that the dynamics will be driven by evolution in two zones of size $\epsilon$ near the boundaries $x = 0$ and $x = 1$. We begin with the fully mixed system in \S\ref{sec:fullymixed}, and generalize to the class-based system in \S\ref{sec:cbmf_numeric_approx}.

\subsection{Fully mixed model}\label{sec:fullymixed}
We consider the fully mixed mean-field equation~\eqref{eqn:omf} for $\epsilon \ll 1$ and perform the rescalings
\begin{equation}\label{eq:rescaling1}
x = \epsilon\tilde{x}, \quad z = \epsilon\tilde{z}, \quad t = \tau/\epsilon, \quad \tilde{P}(\tilde{x},\tau) = P(x,t)
\end{equation}
and
\begin{equation}\label{eq:rescaling2}
    x = 1-\epsilon\tilde{x}, \quad z = -\epsilon \tilde{z}, \quad t = \tau/\epsilon, \quad \tilde{P}(\tilde{x},\tau) = P(x,t)
\end{equation}
near $x = 0$ and $x = 1$ respectively.  We note that the natural time scaling $t \sim 1/\epsilon$ is required to balance terms in the governing equations~\eqref{eq:chapter2} and~\eqref{eqn:omf}, and reflects the fact that each individual has fewer interactions per unit time as $\epsilon$ decreases. With the rescaling, {the two halves of the symmetric} opinion space $x\in[0,1]$ {are each} mapped onto the space $\tilde{x}\in[0,1/(2\epsilon)]$. Both boundaries of the $x$-space, $x=0$ and $x=1$, are mapped into the left boundary $\tilde{x}=0$ while the right boundary $\tilde{x}=1/(2\epsilon)$ corresponds to the centre of the $x$-domain (see Fig.\ref{fig:regions_new}b). In the new space, the disturbance to the initial distribution comes from the left boundary and gradually propagates to the rest of the domain as time advances. Moreover, the right boundary $\tilde{x}={1/(2\epsilon)}$ goes to infinity as $\epsilon \to 0$. Substituting the rescalings~\eqref{eq:rescaling1}---\eqref{eq:rescaling2} into the Ben-Naim~{\etal~equation}~\eqref{eqn:omf} we obtain a new system of governing equations
\begin{equation}
\label{eqn:rescaledsingleclass}
  \frac{\partial \tilde{P}}{\partial \tau}  =
  \left\{
  \begin{array}{cc}
  \displaystyle
  \int_{-2\tilde{x}}^{2\tilde{x}}\tilde{P}\left(\tilde{x}+\frac{\tilde{z}}{2},\tau\right)\tilde{P}\left(\tilde{x}-\frac{\tilde{z}}{2},\tau\right)\,d\tilde{z}
    -\int_{-\tilde{x}}^{1}\tilde{P}(\tilde{x},\tau)\tilde{P}(\tilde{x}+\tilde{z},\tau)\,d\tilde{z},\phantom{sp} &  0 \leq \tilde{x} \leq \frac{1}{2} \\
    \displaystyle
  \int_{-1}^{1}\tilde{P}\left(\tilde{x}+\frac{\tilde{z}}{2},\tau\right)\tilde{P}\left(\tilde{x}-\frac{\tilde{z}}{2},\tau\right)\,d\tilde{z} 
    -\int_{-\tilde{x}}^{1}\tilde{P}(\tilde{x},\tau)\tilde{P}(\tilde{x}+\tilde{z},\tau)\,d\tilde{z}, &  \frac{1}{2} \leq \tilde{x} \leq 1 \\
    \displaystyle
    \int_{-1}^{1}  \tilde{P}\left(\tilde{x}+\frac{\tilde{z}}{2},\tau\right)\tilde{P}\left(\tilde{x}-\frac{\tilde{z}}{2},\tau\right) \,d\tilde{z} -  \int_{-1}^{1} \tilde{P}(\tilde{x},\tau)\tilde{P}(\tilde{x}+\tilde{z},\tau)\,d\tilde{z}, & 1<\tilde{x} < \infty, 
    \end{array}
    \right.
    \end{equation}
with initial condition $\tilde{P}(\tilde{x},0) = 1$.  
In the far-field $\tilde{x} \gg 1$, we can Taylor-expand the {integrands} in~\eqref{eqn:rescaledsingleclass} to find $\partial \tilde{P}/\partial \tau \approx 0$ and hence $\tilde{P}(\tilde{x}) \to 1$ as $\tilde{x} \to \infty$.  The rescaled problem~\eqref{eqn:rescaledsingleclass} must be solved numerically, but has no parameter dependence, and the solution gives an approximation of the probability distribution for any $\epsilon \ll 1$, at any $\tau \ll 1/\epsilon$.

\begin{figure}[ht]
\centering
\includegraphics[width=15cm]{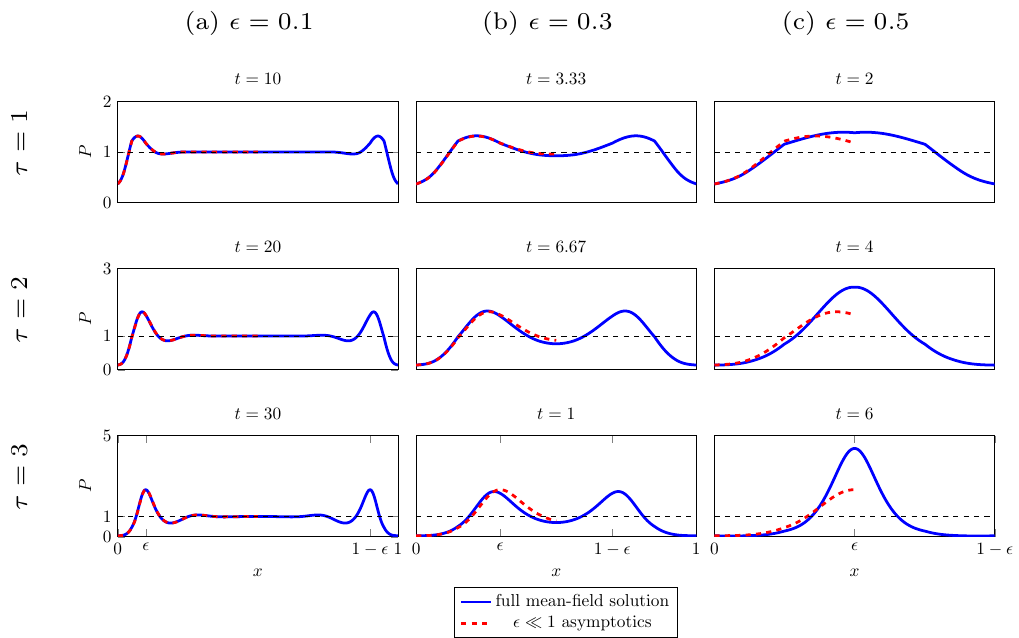}
\caption{Numerical solution to the fully-mixed mean-field model~\eqref{eqn:omf}, plotted together with the small-$\epsilon$ asymptotic approximation determined by solving {the canonical problem}~\eqref{eqn:rescaledsingleclass}.  The initial distribution $P(x,0)=1$ is included as a dashed horizontal line on all panels for reference. The asymptotic approximation holds on a domain $\tilde{x} \in [0,\infty)$ as illustrated in Fig.~\ref{fig:regions_new}; here it is illustrated on the left-hand side of the domain only and truncated at $\epsilon \tilde{x} = 0.5$.    \label{fig:figure3new}}
\end{figure}

The numerical solution to the canonical  problem~\eqref{eqn:rescaledsingleclass} gives a good approximation to the $\epsilon = 0.1$ solution of the full problem~\eqref{eqn:omf} (see Fig.~\ref{fig:figure3new}a).  We expect this solution to break down at $\tau = O(1/\epsilon)$ when the far-field condition ceases to be valid.  
For $\epsilon = 0.3$ and $\epsilon = 0.5$, the accuracy of the approximation is limited by the fact that the two boundary regions near $x = 0$ and $x = 1$ are not independent (see Fig.~\ref{fig:figure3new}b,c).  For $\epsilon = 0.5$, the system forms a single cluster near $x = 0.5$, which cannot be resolved within the framework of the asymptotic approximation $P \to 1$ as $x/\epsilon \to \infty$ and/or $(1-x)/\epsilon \to \infty$. 

\begin{figure}[ht]
    \centering
    \begin{tikzpicture}
    
    \node at (0,0) {\includegraphics[width=7.0cm]{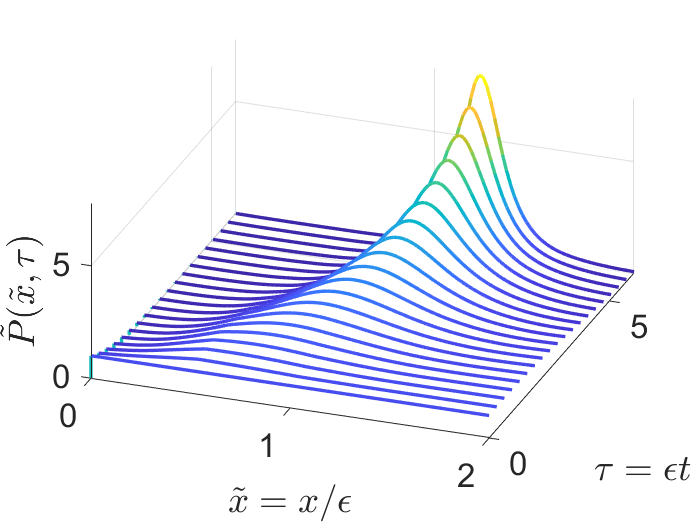}};
    \node at (8.5,0) {\includegraphics[width=7.0cm]{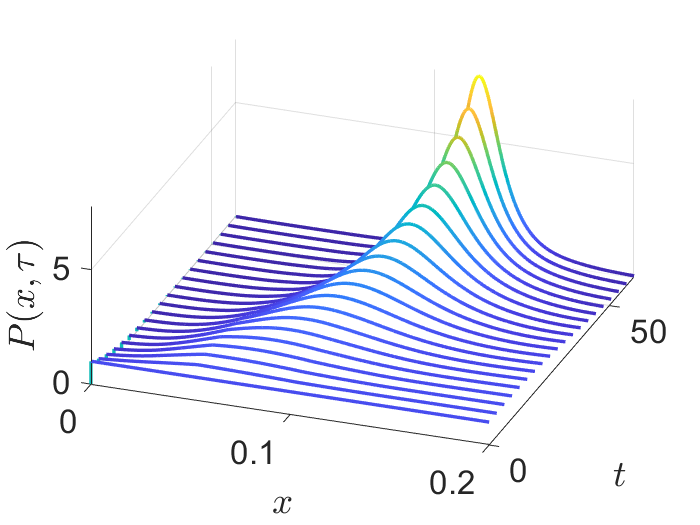}};
    \node at (-3,2) {(a)};
    \node at (5.5,2) {(b)};
    \end{tikzpicture}
\caption{\label{fig:figure4new} Mass in the boundary region near $x=0$ as predicted by (a) the $\epsilon \ll 1$ approximation and (b) the full solution to the {mean-field} equation for $\epsilon = 0.1$. }
\end{figure}

We gain further insight by tracking the motion of mass in the system as it evolves (see Fig.~\ref{fig:figure4new}).  In the asymptotic approximation, mass moves out of the interval $\tilde{x} < 1$ and towards a cluster at approximately $\tilde{x} = 1.2$ (Fig.~\ref{fig:figure4new}a).  As $\tau$ increases, the mass in the boundary region tends to zero.  This is consistent with the prediction from numerical solution of the full mean-field equation (Fig.~\ref{fig:figure4new}b).

\begin{figure}[ht]
\centering
\begin{tikzpicture}
\node at (0,0) {\includegraphics[trim = 0mm 20mm 0mm 0mm, clip, width=8cm]{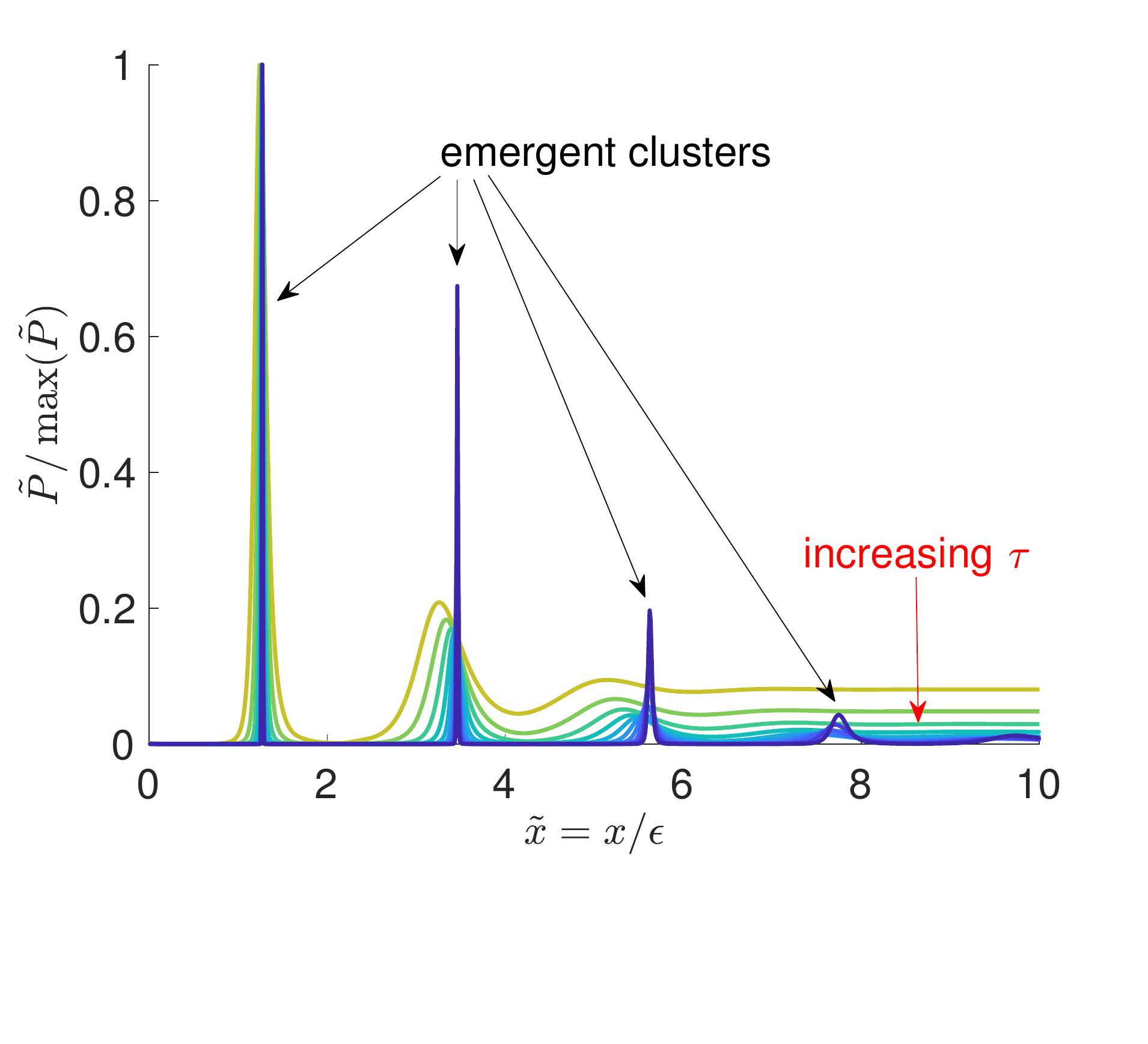}};
\node at (8,0) {\includegraphics[trim = 0mm 20mm 0mm 0mm, clip, width=8cm]{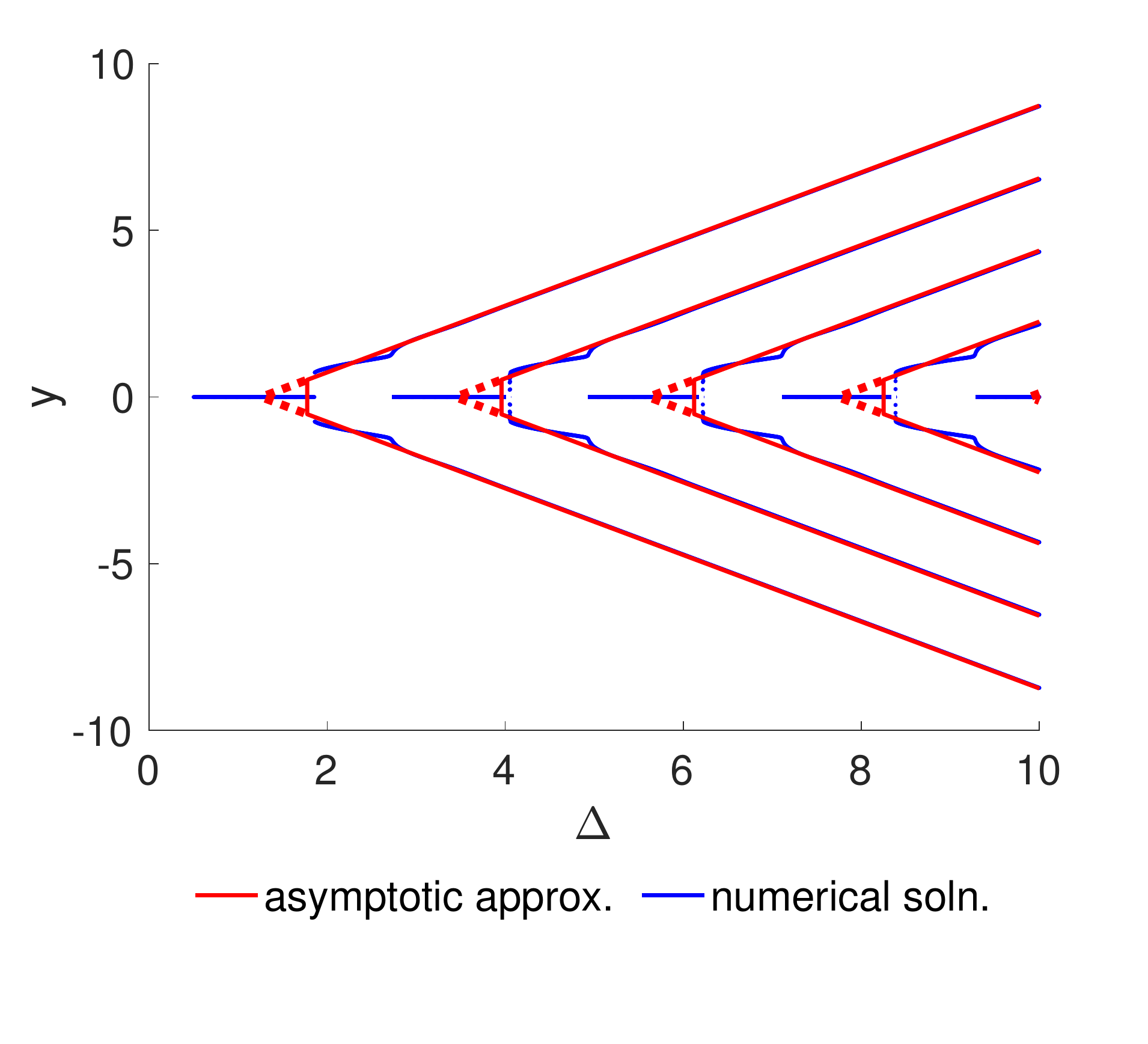} };
\node at (-4,3) {(a)};
\node at (4,3) {(b)};
\end{tikzpicture}
{\caption{(a) Normalized plot of {solution $\tilde{P}$ of the canonical problem~\eqref{eqn:rescaledsingleclass}}  at $\tau = 8,9,\cdots,16$, highlighting the emergence of clusters at a regular {$\tilde{x}$}  spacing of approximately 2. (b) The positions of the opinion clusters reconstructed from $\tilde{P}$ peaks of (a) (red) and numerical solution of the full mean-field equation (blue). The red dotted lines show the {symmetric} clusters which cannot exist in the $t\to\infty$ limit  as they are separated by a distance less than $\epsilon$ (equivalent to $1$ in $(y,\Delta)$ space). Instead, they should merge into a single cluster located {at the midpoint}. } \label{fig:clustersandbifurcation}}
\end{figure}

Next, we examine the late-time behaviour of the solution of the canonical problem ~\eqref{eqn:rescaledsingleclass}. We observe that peaks emerge at a regular spacing of approximately $2$ as $\tau$ increases (see Fig.~\ref{fig:clustersandbifurcation}a). These peaks form at positions 
\begin{equation}\label{eq:canonical_peaks}
    \tilde{x}^*=\{1.26,3.44,5.61,7.74,9.89 \dots\},
\end{equation}
and are associated with the emergence of opinion clusters. Note that the mass gets locked in the emergent clusters at quite early times. This is due to the fact that the clusters emerge with separation greater than the confidence bound and cannot interact thereafter.
At later times, the mass locked in a cluster moves into a delta-function-like peak, and while the locations of the delta peaks can differ from the centre of the early-time peaks, this difference is small. Therefore, we can obtain a good approximation to the locations of the opinion clusters in the infinite-time limit solution from the early-time peaks $\tilde{x}^*$. We perform this by selecting $\tilde{x}^*$ that are within the $0\leq \tilde{x}^*\leq\Delta$ interval and applying the reverse transformation of \eqref{eq:rescaling1}---\eqref{eq:rescaling2}. 

In Figure~\ref{fig:clustersandbifurcation}b, we show the predicted positions of final clusters in $(y,\epsilon)$ space.  Combining the transformations~\eqref{eq:rescaling1}, \eqref{eq:rescaling2} and \eqref{eq:transformation} with the result~\eqref{eq:canonical_peaks}, we obtain
\begin{equation}
    y^* = \pm \left( \tilde{x}^* - \frac{1}{2\epsilon} \right) = \pm \left( \tilde{x}^* - \Delta \right).
\end{equation}
Comparing with the solution of the full mean-field problem~\eqref{eqn:omf}, 
we see that the asymptotic approximation predicts the location of symmetric clusters very accurately. However, the central clusters do not appear in this approximation. Instead, we observe two symmetric clusters situated close to the domain centre within the distance less than the confidence bound from each other. These clusters obviously cannot exist in the infinite-time limit solution, but should merge into a single cluster located at the centre. 
The failure of the current analysis to predict central clusters is {a consequence of the assumption} $P \to 1$ as $\tilde{x} \to \infty$, which is clearly untrue when there is a central cluster. These limitations will be addressed by a linear stability analysis in \S\ref{sec:linearstability}.

\subsection{Class-based model} \label{sec:cbmf_numeric_approx}
We now extend the analysis from Sec.~\ref{sec:fullymixed} to  the class-based model from~\cite{Fennell2021}. Carrying out the same rescaling~\eqref{eq:rescaling1}---\eqref{eq:rescaling2} for the class-based model \eqref{eq:chapter2} from~\cite{Fennell2021} yields
\begin{align}
\label{eq:rescaledclasses1}
    & \frac{\partial \tilde{P}_l}{\partial \tau} 
    = \sum_j C_{lj}\left[\int\limits_{-2\tilde{x}}^{2\tilde{x}}\tilde{P_l}\left(\tilde{x}+\frac{\tilde{z}}{2},\tau\right)\tilde{P_j}\left(\tilde{x}-\frac{\tilde{z}}{2},\tau\right)\,\mathrm{d}\tilde{z}
    -\int\limits_{-\tilde{x}}^{1}\tilde{P_l}(\tilde{x},\tau)\tilde{P_j}(\tilde{x}+\tilde{z},\tau)\,\mathrm{d}\tilde{z}\right],\phantom{sp} &  0 \leq \tilde{x} \leq \frac{1}{2} \\
    \label{eq:rescaledclasses2}
    &\frac{\partial \tilde{P}_l}{\partial \tau} 
    = \sum_j C_{lj}\left[\int\limits_{-1}^{1}\tilde{P_l}(\tilde{x}+\frac{\tilde{z}}{2},\tau)\tilde{P_j}\left(\tilde{x}-\frac{\tilde{z}}{2},\tau\right)\,\mathrm{d}\tilde{z} 
    -\int\limits_{-\tilde{x}}^{1}\tilde{P_l}(\tilde{x},\tau)\tilde{P_j}(\tilde{x}+\tilde{z},\tau)\,\mathrm{d}\tilde{z}\right], \phantom{sp}&  \frac{1}{2} \leq \tilde{x} \leq 1 \\
    \label{eq:rescaledclasses3}
    &\frac{\partial \tilde{P_{l}}}{\partial \tau} =  \sum_j C_{lj}\left[\int\limits_{-1}^{1}\tilde{P_l}(\tilde{x}+\frac{\tilde{z}}{2},\tau)\tilde{P_j}\left(\tilde{x}-\frac{\tilde{z}}{2},\tau\right)\,\mathrm{d}\tilde{z}
    -\int\limits_{-1}^{1}\tilde{P_l}(\tilde{x},\tau)\tilde{P_j}(\tilde{x}+\tilde{z},\tau)\,\mathrm{d}\tilde{z}\right], & 1 < x < \infty, 
   \end{align}
for each class $l$ in the network, with $\tilde{P_l}(\tilde{x},0) = 1$ and $\tilde{P}_l \to 1$ as $\tilde{x} \to \infty$. 
As an example, we consider a network with two degree classes, where $90\%$ of nodes have degree $k_{maj}$ and $10\%$ of nodes have degree 100, as studied in~\cite{Fennell2021}. 

\begin{figure}
\centering
\begin{tikzpicture}
\node at (0,0){\includegraphics[width=14cm]{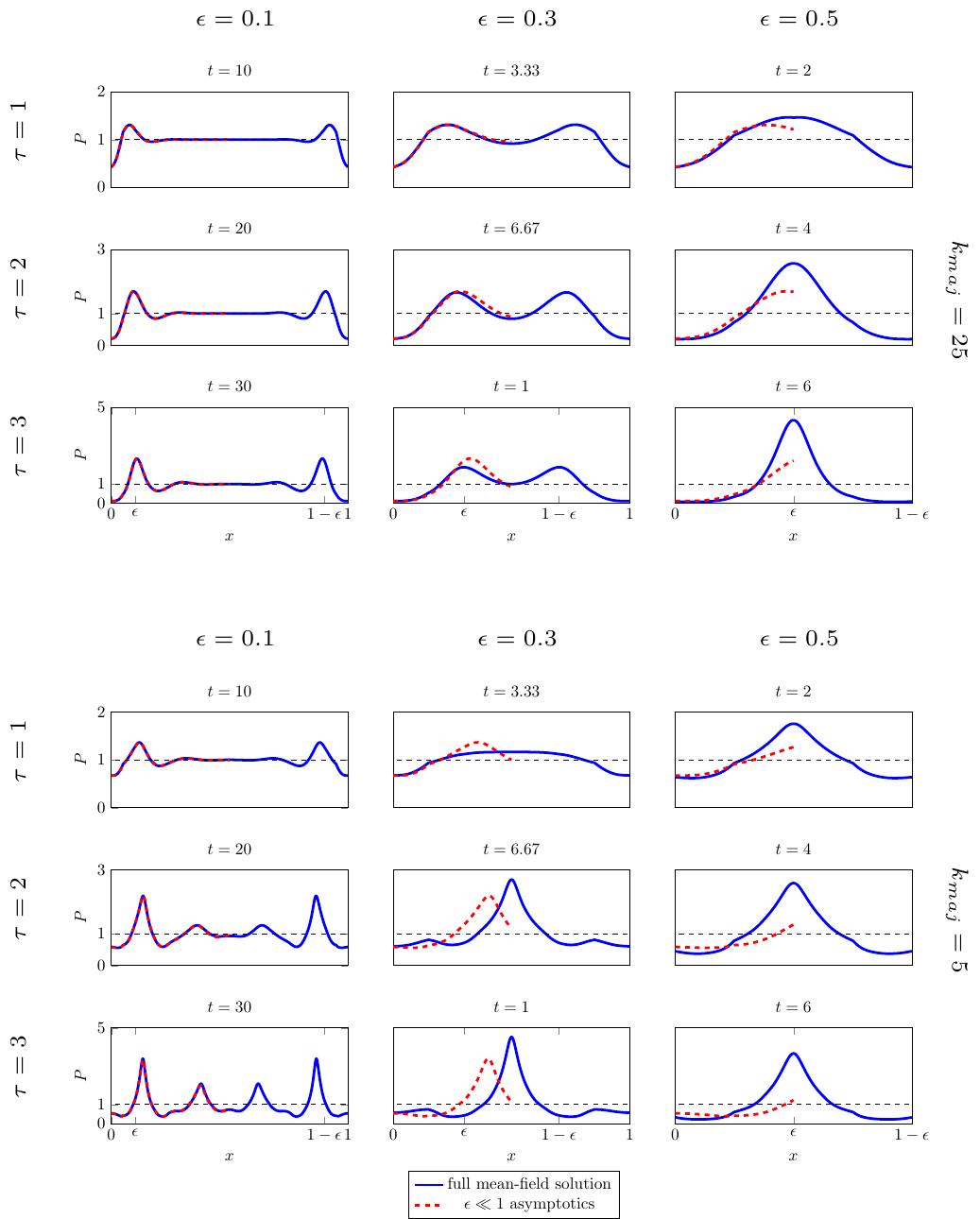}};
\node at (-7.2,8.5) {(a)};
\node at (-7.2,-0.4) {(b)};
\end{tikzpicture}
{\caption{The numerical solution and small-$\epsilon$ approximation to the opinion distribution for a network in which $90\%$ of nodes have degree $k_{maj}$ and $10\%$ of nodes have degree 100, with $k_{maj} = 25$ (top panels) and $k_{maj} = 5$ (bottom panels). The initial distribution $P(x,0)=1$ is included {as a dashed horizontal line} on all panels for reference.} \label{fig:figure5new}}
\end{figure}

Figure~\ref{fig:figure5new} shows the opinion distribution across the whole population for the cases $k_{maj} = 25$ and $k_{maj}=5$. The $\epsilon \to 0$ approximation is in excellent agreement with the full mean-field solution for $\epsilon = 0.1$, as expected.  For epsilon = $0.3$ the model gives a reasonable estimate of the distribution for $k_{maj} = 25$, while for $\epsilon = 0.5$ the agreement is poor, as expected for our small-$\epsilon$ approximation.  When $k_{maj} = 5$, the probability distributions are noticeably different, and so too is the performance of the asymptotics.

\begin{figure}[ht]
    \centering
    \includegraphics[width=13.8cm]{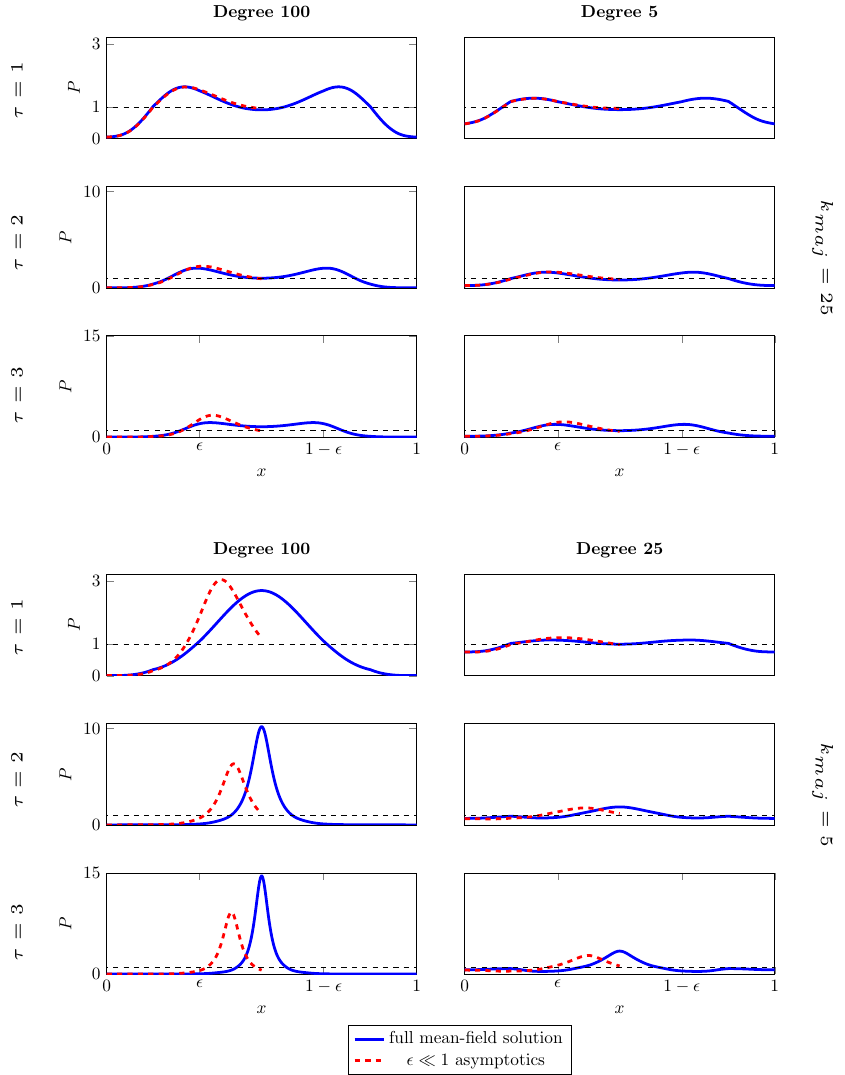}
    \caption{The distribution within each degree class for confidence bound $\epsilon = 0.3$.  In the $k_{maj}=25$ case (top panels), the distributions for the degree-100 and degree-25 nodes evolve on comparable timescales.  In the $k_{maj} = 5$ case (bottom panels), the distribution for the degree-100 nodes evolves much more quickly than the distribution for the degree-5 nodes. The initial distribution $P(x,0)=1$ is included as a dashed horizontal lines for reference.}
    \label{fig:figure7new}
\end{figure}

In the case where $k_{maj} = 5$, $90\%$ of nodes are very weakly connected, with degree $5$, and the associated opinions evolve very slowly compared to the well-connected $10\%$ (with degree $100$), due to the disparate coefficients $C \approx (0.1, 0.2; 2.1, 4.8)$ in the evolution equation~\eqref{eq:chapter2}. The opinions of the small but well-connected $10\%$ evolve relatively quickly, but are influenced by interactions with an almost stationary community where $P \approx 1$ $\forall$ $x \in [0,1]$ and so they are pulled towards the centre of opinion space.  We observe (Fig.~\ref{fig:figure7new}) that in the system where $k_{maj} = 5$, $P$ for the degree-100 group is already close to 3 at $x = 0.5$ when $\tau = 3$, while $P \approx 1$ in the degree-5 group at the same time.  This is in contrast to the $k_{maj} = 25$ scenario, where $C \approx (0.5, 0.2; 2.1, 0.9)$ and the opinions of the two classes evolve at more similar speeds.  In the $k_{maj} = 5$ case, where the two classes evolve on disparate timescales, the increased mass of well-connected nodes near $x = 0.5$ acts to attract the less connected (and therefore more slowly evolving) group towards this centre, and a small grouping emerges there by $\tau = 3$.  
These compounding effects lead to an increase in the extent to which mass is drawn towards the centre of opinion space in the case where the degrees of the two classes are very different (Fig.~\ref{fig:figure5new}b) compared to the case where the two degree classes are less distinct (Fig.~\ref{fig:figure5new}a) or where there is only one degree class (Fig.~\ref{fig:figure3new}).
Since the asymptotic analysis relies on the existence of two independent boundary regions with $P(x\to\infty,t) = 1$, it does not accurately predict these dynamics even for $\epsilon = 0.3$.

\begin{figure}[ht]
\centering
\begin{tikzpicture}
\node at (0,0){\includegraphics[width=14cm]{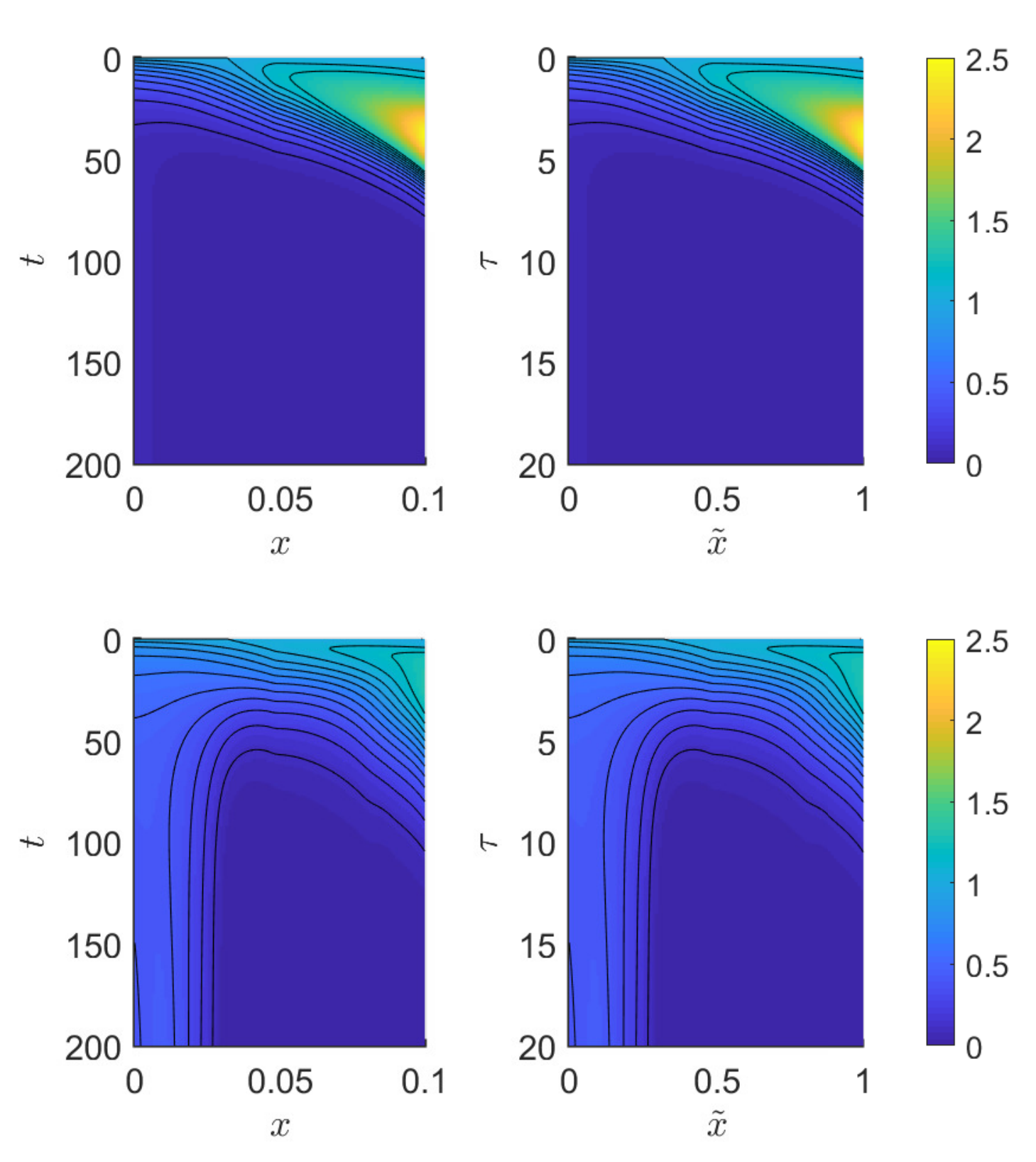}};
\node at (-3.2,8) {$\epsilon = 0.1$ solution};
\node at (2.9,8) {$\epsilon \to 0$ limit};
\node[rotate=90] at (-7.8,4) {$k_{maj} = 25$};
\node[rotate=90] at (-7.8,-3.5) {$k_{maj} = 5$};
\end{tikzpicture}
{\caption{Mass in the boundary region near $x = \tilde{x} = 0$ for a network in which $90\%$ of nodes have degree $k_{maj}$ and $10\%$ of nodes have degree 100. The numerical solution to the full problem~\eqref{eq:chapter2} for $\epsilon = 0.1$ is shown (left) together with the numerical solution for the $\epsilon \to 0$ limit~\eqref{eq:rescaledclasses1}---\eqref{eq:rescaledclasses3}, for the cases $k_{maj} = 25$ (top) and $k_{maj} = 5$ (bottom).  Contour lines are included between 0.1 and 1.2 at intervals of 0.1.  All mass flows to the central region (towards the right) for $k_{maj}=25$, while some mass persists close to $x=0$ for $k_{maj}=5$.} \label{fig:figure6new}}
\end{figure}

Finally, we consider in Fig.~\ref{fig:figure6new} the question of whether some agents retain ``extreme'' opinions near $x = 0$ and $x = 1$, even as time increases.  This phenomenon was observed previously for the $k_{maj}=5$ case in~\cite{Fennell2021}, and is reproduced here for the $\epsilon = 0.1$ solution of the full problem~\eqref{eq:chapter2} as well as the $\epsilon \to 0$ limit~\eqref{eq:rescaledclasses1}---\eqref{eq:rescaledclasses3}.  This verifies that the phenomenon exists for all $\epsilon \ll 1$, provided there is sufficient contrast between the degrees of the two classes.

\subsection{{Implications}}
The regime $\epsilon \ll 1$ describes situations where the confidence bound on a network is small.  Our analysis demonstrates that opinions evolve slowly when interactions {occur only} over a limited opinion range.  If the opinion density is initially uniform {(as is usually assumed in the literature)}, then the density of opinions in the centre of opinion space is fixed at leading order, and opinion evolution is driven by an imbalance in interactions near the boundaries.  These interactions and the consequent opinion evolution are repeated across different values of $\epsilon$ over an opinion distance that scales with $\epsilon$.  This analysis results in a canonical boundary-region problem that must be solved numerically but is independent of the confidence bound $\epsilon$. For a two-class case study, we have demonstrated that the proportion of the population left in the boundary regions depends on the degree of the majority {of nodes}, in line with the results {obtained numerically in}~\cite{Fennell2021}.  For the one-class case, our asymptotic solution points to the emergence of clusters at late times (Fig.~\ref{fig:clustersandbifurcation}a), but cannot resolve the emergence of these clusters across the whole domain.  We require a linear stability analysis that is not predicated on $\epsilon \to 0$ to properly predict opinion clusters; this is the topic of \S\ref{sec:linearstability}. 

\section{Cluster location: linear stability analysis}\label{sec:linearstability}
We saw in Sec.~\ref{sec:meanfield} that the distribution in the central region evolves more slowly than the distribution in the boundary regions at early times. This is illustrated by the numerical solutions (see, e.g.,\ Figures~\ref{fig:figure3new} and \ref{fig:figure5new}), and quantified by  $\partial P/\partial \tau = O(\epsilon)$ in the central region in Equations~\eqref{eqn:rescaledsingleclass} and~\eqref{eq:rescaledclasses1}---\eqref{eq:rescaledclasses3}.  
In fact, if there were no boundaries then $P \equiv 1$ would be a steady-state solution to the fully-mixed mean-field equation~\eqref{eqn:omf} (or equivalently $P_k\equiv 1$ would be a steady-state solution to the class-based mean-field equation~\eqref{eq:chapter2}).
The initial deviation from $P \equiv 1$ is driven by an imbalance in influence at the boundaries, but we hypothesize here that the number of emergent clusters is not affected by boundaries.  We neglect the role of the boundaries in this section and consider a linear stability analysis~\cite{Logan2013} about a uniform steady state in order to assess the response of the system to a fluctuation in opinions.  By identifying the fastest-growing mode, we estimate the number of opinion clusters that form across opinion space.  We first consider the single-class case in \S\ref{sec:omf_linear_stability}---\ref{sec:newsub2} and then generalize to multiple degree classes in \S\ref{sec:cbmf_lin_stability}---\ref{sec:twoclasslinstab}.

\subsection{Small perturbation}\label{sec:omf_linear_stability}
We start by rewriting~\eqref{eqn:omf} in a format more amenable to linear stability analysis.  We assume that $P(x,t)$ is analytic everywhere, and rewrite the integrands in terms of their Taylor expansions.  Evaluating the integrals then yields
\begin{multline}
\label{eqn:equation_in_central_region}
\frac{\partial P(x,t)}{\partial t} 
=  \sum_{m=1}^{\infty}2\epsilon^{2m+1}\left[ P(x,t)P^{(2m)}(x,t)\left(\frac{(1/2)^{2m-1} -1}{(2m+1)!}\right)
\right.\\
+
\left.
\sum_{n=1}^{2m-1}P^{(2m-n)}(x,t)P^{(n)}(x,t)\frac{(1/2)^{2m}(-1)^n }{(2m-n)!n!(2m+1)} \right],
\end{multline}
where $P^{(n)}(x,t)$ denotes the $n$th derivative of $P$ with respect to $x$. We note here that the $(2m)$ exponent arises from the fact that only even functions contribute to the integral, that the $m=0$ contribution to the sum is eliminated by cancellation between $I_1$ and $I_2$, and that the $n=0$ and $n=2m$ contributions to the second sum have been incorporated into the first term on the right-hand side of~\eqref{eqn:equation_in_central_region}.  We also emphasize that the form of~\eqref{eqn:equation_in_central_region} is not dependent on the size of $\epsilon$ --- since all terms in the Taylor series are retained, the equation formally holds as long as the series converges. 

We now consider a perturbation of the form $P(x,t) = 1 + \delta \phi(x,t)$ {for small $\delta$}.  Substituting this into~\eqref{eqn:equation_in_central_region} and linearizing yields
\begin{align}
\label{eqn:linearized_sys}
\frac{\partial \phi(x,t)}{\partial t} 
&= \sum_{m=1}^{\infty}2\epsilon^{2m+1}\left[\phi^{(2m)}(x,t)\left(\frac{(1/2)^{2m-1} -1}{(2m+1)!}\right)\right],
\end{align}
with initial perturbation $\phi(x,0)$. In the original fully-mixed mean-field equation~\eqref{eqn:omf}, the change in the distribution in the central region is driven by the change in the solution in the boundary regions. In the linear stability analysis the change in the distribution is driven by the initial perturbation $\phi(x,0)$. We therefore focus in the analysis below on the central region $[\epsilon, 1-\epsilon]$.
Since the full solution is symmetric, we also assume the initial perturbation $\phi(x,0)$ is symmetric. In the linear stability analysis the initial perturbation is assumed to be small, and therefore this approach is only valid when the ``perturbation'' from the boundary regions in the full solution is small. This happens when the mass in the boundary regions is small relative to the mass in the central region, i.e., when the central region is wider than the boundary {layer}.

\subsection{Modal decomposition}
\begin{figure}[ht]
    \centering
        \centering
        \includegraphics[width=0.4\textwidth]{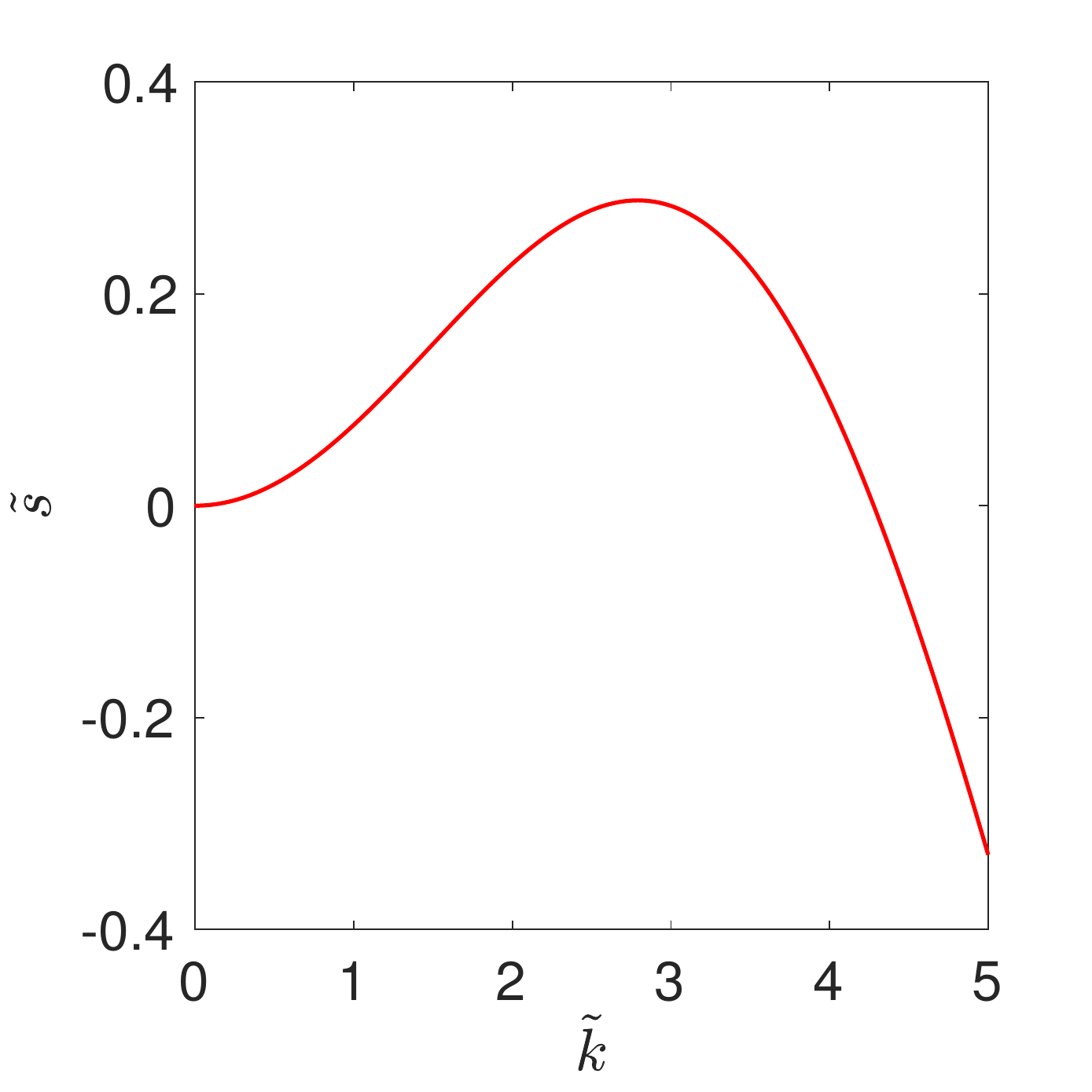} 
        \caption{\label{fig:s_vs_x}The growth rate as a function of $\tilde{k} = \epsilon k$.  We only consider $\tilde{k} \in [0,5]$ since the growth rate $\tilde{s}(\tilde{k})$ is negative for $\tilde{k}>5$.}
\end{figure}

Since the full solution is symmetric we assume symmetric boundary conditions $\phi(\epsilon,t) = \phi(1-\epsilon,t)$. 
A function on a finite domain with symmetric boundary conditions can be written as a Fourier series, and therefore Eq.~\ref{eqn:linearized_sys} has fundamental solutions $A(t)e^{ikx}$ where $k={2n\pi}/{L}$ for integer $n$ and $L=1-2\epsilon$ is the width of the central region. Plugging $\phi(x,t) = A(t)e^{ikx}$ into Eq.~\ref{eqn:linearized_sys} and evaluating the resulting sums yields an evolution equation
\begin{equation}
\frac{\mathrm{d}A(t)}{\mathrm{d}t}
=\left(\frac{8\sin(\epsilon k/2 )}{k} -\frac{2\sin(\epsilon  k)}{k} -2\epsilon\right)A(t),
\end{equation}
which has solution $A(t) = e^{st}$ {with} 
\begin{equation}
s = s(k,\epsilon)=\frac{8\sin(\epsilon k/2 )}{k} -\frac{2\sin(\epsilon  k)}{k} -2\epsilon.
\end{equation}
The general solution to the linearized system is thus
\begin{equation*}
\phi(x,t) = \sum_{n=-\infty}^{\infty}{c}_ne^{s_nt+ik_nx},
\end{equation*}
where $k_n = {2n\pi}/({1-2\epsilon})$, $s_n = s(k_n,\epsilon)$ and the coefficients {$c_n$} depend on the initial perturbation $\phi(x,0)$. The perturbation $\phi(x,t)$ will grow if there are values of $n$ for which $s_n>0$. The solution will be dominated by the fastest growing mode $e^{ik_{n^*}x}$, which is the mode which has the largest positive growth rate $s_{n^*}$. To find the fastest growing mode we must find the maximum value of $s$, which will vary depending on $\epsilon$. If we write $\tilde{k}=k\epsilon$ and $\tilde{s}(\tilde{k}) = s(k,\epsilon)/(2\epsilon)$ then we have an equation which is independent of $\epsilon$,
\begin{equation}\label{eq:s(k)}
\tilde{s}(\tilde{k}) = \frac{4\sin\left(\frac{\tilde{k}}{2}\right)}{\tilde{k}} - \frac{\sin\left(\tilde{k}\right)}{\tilde{k}}-1.
\end{equation}
We note that this is the same as the dispersion relation obtained in~\cite{pineda2009} via linear stability analysis of the Deffuant model with periodic boundary condition (i.e. with the opinion space wrapped on a circle), if noise is neglected in that study.
The growth rate $\tilde{s}(\tilde{k})$ is maximized at 
$\tilde{k}^* = 2.79$ (see Fig.~\ref{fig:s_vs_x}), and so $s$ {is} maximized when $k = 2.79/\epsilon$. In general, however, this value of $k$ will not be of the form $k = {2n\pi}/{(1-2\epsilon)}$ for integer $n$. To find the fastest growing mode we determine the value of $n\in \mathbb{Z}$ for which ${2n\pi}/({1-2\epsilon})\leq k^*<{2(n+1)\pi}/{{(1-2\epsilon)}}$, and then select whichever of ${2n\pi}/({1-2\epsilon})$ or ${2(n+1)\pi}/({1-2\epsilon})$ corresponds to a larger value of $s$.

\noindent

\subsection{Clusters}\label{sec:newsub2}
\begin{figure}[ht]
\centering
\begin{tikzpicture}
\node at (0,0) {\includegraphics[trim = 0mm 20mm 0mm 0mm, clip, width=7.6cm]{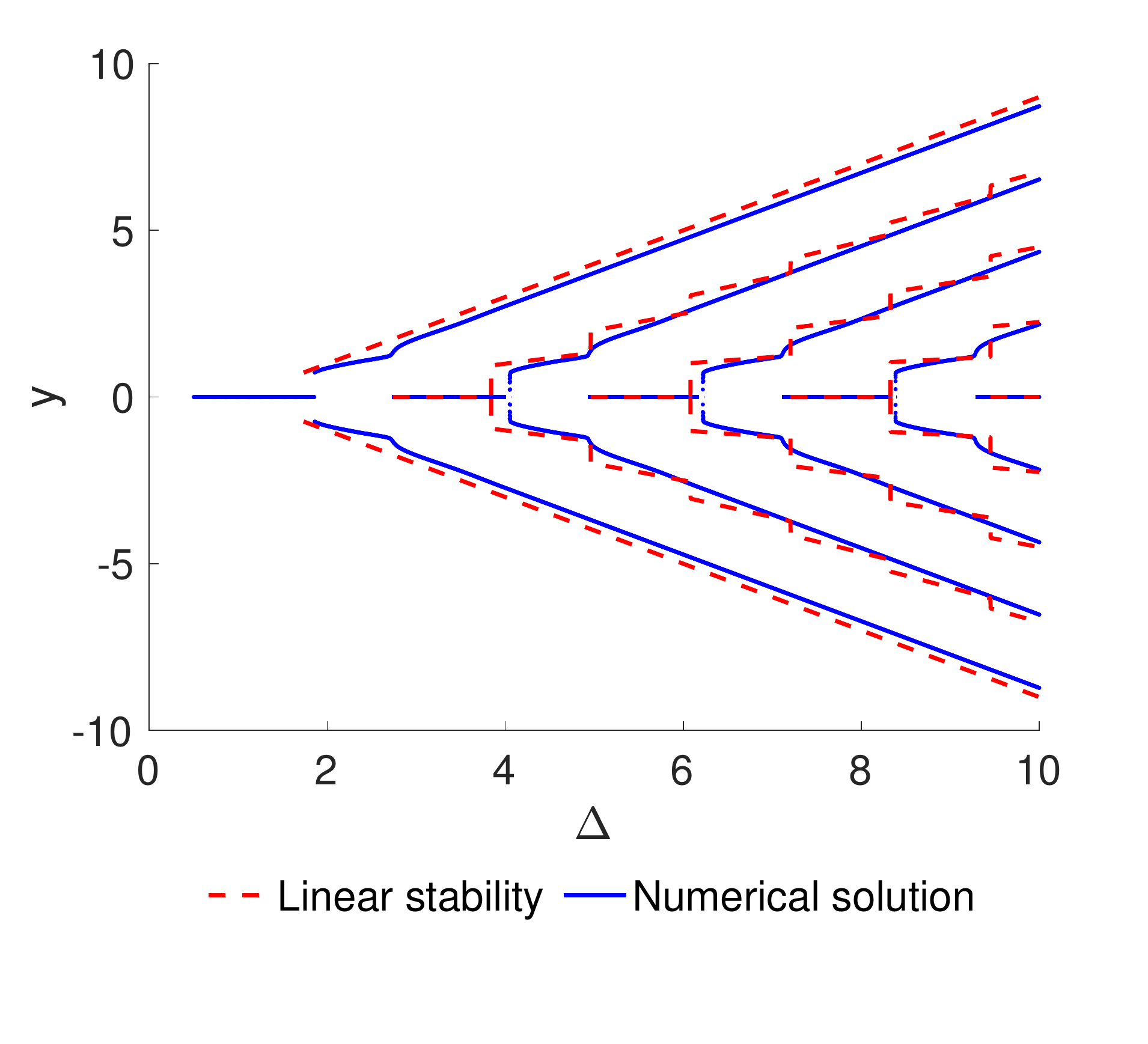}};
\node at (8.2,0) {\includegraphics[trim = 0mm 20mm 0mm 0mm, clip, width=7.6cm]{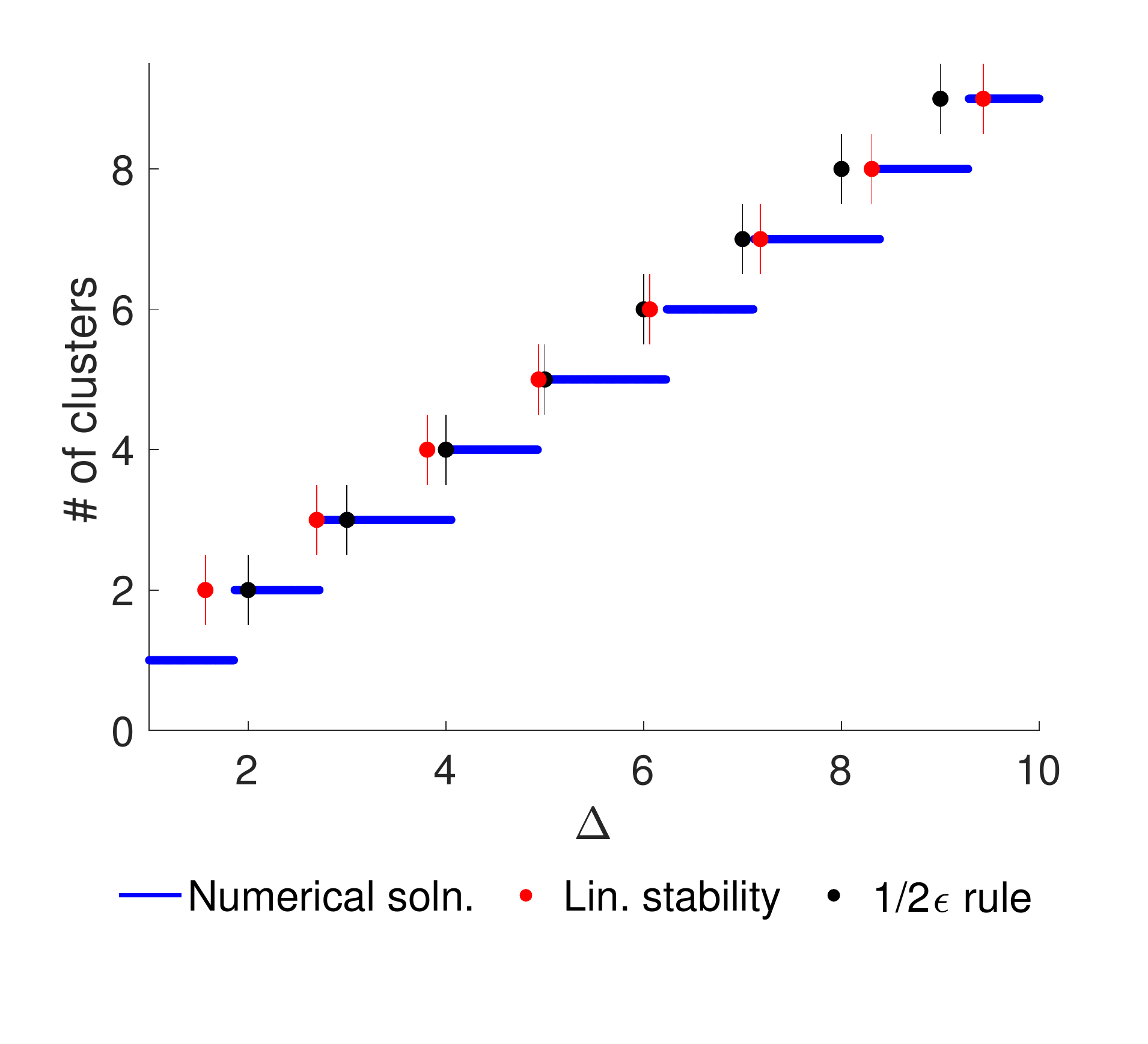}};
\node at (-4,3) {(a)};
\node at (4.3,3) {(b)};
\end{tikzpicture}
\caption{\label{fig:bif_diag_delta} (a) Location of peaks predicted by the linear stability analysis (red) and numerics (blue) for the original mean-field equation in  $(y,\Delta)$ space. (b) The number of opinion clusters in the final opinion configuration from numerical simulations (blue lines), predicted by Eq.~\eqref{eq:n_clusters} (red dots) and the $1/(2\epsilon)$ rule of~\cite{Ben-Naim2003} (black dots).}
\end{figure}

The fastest-growing mode identified above gives a prediction for the number of clusters that should be expected on the opinion space $[0,1]$. We note that a natural rescaling $k^{-1} \sim \epsilon$ was used to derive~\eqref{eq:s(k)}.  This indicates that the distance between clusters should scale linearly with the confidence bound $\epsilon$, and is consistent with~\cite{Ben-Naim2003}, where the authors suggest that the number of clusters on an interval $[0,1]$ is given by the integer part of $1/(2\epsilon)$.  
We can go a step further here using the results of our linear-stability analysis. 
For a symmetric initial perturbation $\phi(x,0)$ the fastest-growing mode is $\cos\left(\frac{2n^*\pi}{1-2\epsilon} (x-\epsilon)\right)$. If we assume peaks form at the boundaries  $x=\epsilon$ and $x=1-\epsilon$ of the central region, which is consistent with our numerical solutions from \S\ref{sec:omf_numerical_approx}, then the peaks of this function are at
\begin{equation}\label{eq:cluster_pos}
    x_m=\epsilon+\frac{m(1-2\epsilon)}{n^*},\quad m \in \{0,\cdots,n^*\},
\end{equation}
where $n^*$ is given by $\frac{2.79}{\epsilon} \approx\frac{2n^*\pi}{1-2\epsilon}$, i.e., ${n^* \approx\frac{2.79}{\pi}\left(\frac{1}{2\epsilon} - 1\right)}$. 
These are the points in the opinion space where the perturbation is growing the fastest, at least initially, and so this is where we predict the final clusters will be located. 
There will be $n_c = n^* + 1$ clusters in steady state. An approximate equation for the number of clusters as a function of $\epsilon$ predicted from our analysis can be then {expressed} as
\begin{equation}\label{eq:n_clusters}
    n_c \approx {0.89}\dfrac{1}{2\epsilon} + 0.11.
\end{equation}
This is remarkably similar to the estimation $n_c=0.44/\epsilon$ of~\cite{pineda2009} for the Deffuant model on a circle. Formula \eqref{eq:n_clusters} gives a better approximation to the number of clusters observed in numerical simulations than the heuristic $1/(2\epsilon)$ rule of~\cite{Ben-Naim2003}, as Fig.~\ref{fig:bif_diag_delta}b illustrates. The accuracy of our prediction increases as $\epsilon$ decreases ($\Delta$ increases), while the $1/(2\epsilon)$ rule gets noticeably inaccurate for small $\epsilon$ (large $\Delta$). 

The linear stability analysis not only predicts the number of clusters, it also predicts final positions of these clusters for any given $\epsilon$ or $\Delta$. We 
compare our predictions against numerical solutions of the full mean-field equation \eqref{eqn:omf} in Fig.~\ref{fig:bif_diag_delta}a.
It can be seen that the linear stability analysis accurately predicts locations of both symmetric and central clusters for  $\Delta>1.73$ ($\epsilon<0.29$). Below this value there are no unstable modes, i.e. the linear stability analysis predicts that $P(x,t)=1$ is a stable solution, while the numerical solution shows that one cluster forms. 

Since our analysis relies on an assumption of a steady state $P\equiv1$ that is not affected by the boundaries, it is not surprising that the solution breaks down when this is not valid. Strictly speaking, the linear stability analysis {only} gives us information about emergent clusters at early times when the perturbation $\phi$ is still small. However, as we argued above and as Fig.~\ref{fig:bif_diag_delta}a confirms, the early-time results predict the positions of the clusters in the $t\to\infty$ limit with a great degree of accuracy. Indeed, the solution of $P(x,t)$ at any point is affected only by values in the $\epsilon$-vicinity of $x$. At the same time, we observe that the early clusters are separated by distances of approximately $2\epsilon$. This means that the mass gets divided between clusters at early times and the number of clusters does not change later. With time, the early-time clusters only get narrower as the mass locked in them moves towards their centres. 
At larger values of $\Delta$, we expect the accuracy of the linear stability analysis to be limited by the discrepancy between the growth rates of different clusters illustrated in Fig.~\ref{fig:clustersandbifurcation}a.  In particular, we anticipate that the cluster nearest the boundary may become large before clusters far from the boundary experience a disturbance and begin to grow.

\subsection{Class-based model}\label{sec:cbmf_lin_stability}
We now extend our linear stability analysis to the class-based mean-field equations~\eqref{eq:chapter2}.  We take the same approach as we did for the single-class mean-field equation in \S\ref{sec:omf_linear_stability}.  We neglect the boundaries and boundary regions to treat $P_l\equiv1$ $\forall l$ as a steady-state solution.  We then consider the growth of a small perturbation to the solution in order to predict the cluster locations in the central region.  Using the Taylor series for $P_l(x+z/2,t)$, $P_l(x-z/2,t)$ and $P_l(x+z,t)$ in~\eqref{eq:chapter2} yields
\begin{equation}
\begin{split}
\label{eqn:taylor_cbmf}
&\frac{\partial P_l(x,t)}{\partial t} 
= \sum_j C_{lj}\left( \sum_{m=1}^{\infty}2\epsilon^{2m+1}\left[\sum_{n=1}^{2m-1}P_l^{(2m-n)}(x,t)P_j^{(n)}(x,t)\frac{(1/2)^{2m}(-1)^n }{(2m-n)!n!(2m+1)} \right.\right. \\
&\phantom{space}\left.\left.+ P_l(x,t)P_j^{(2m)}(x,t)\left(\frac{(1/2)^{2m} -1}{(2m+1)!}\right)+P_l^{(2m)}(x,t)P_j(x,t)\left(\frac{(1/2)^{2m}}{(2m+1)!}\right)\right]\right),
\end{split}
\end{equation}
where $P_l^{(n)}(x,t)$ is the $n^{\text{th}}$ derivative {of $P_l$} with respect to $x$.

We consider a small perturbation to the steady state solution for each distribution by writing $P_l(x,t) = 1+\delta \phi_l(x,t)$ for $\delta \ll 1$. Using this in~\eqref{eqn:taylor_cbmf} and linearizing yields
\begin{align}
\label{eqn:lin_approx}
\frac{\partial \phi_l(x,t)}{\partial t} 
= \sum_j C_{lj}&\left( \sum_{m=1}^{\infty}2\epsilon^{2m+1}\left[\phi_j^{(2m)}(x,t)\left(\frac{(1/2)^{2m} -1}{(2m+1)!}\right)
+\phi_l^{(2m)}(x,t)\left(\frac{(1/2)^{2m}}{(2m+1)!}\right)\right]\right).
\end{align}

We assume, as we did in the {single}-class case, that the initial perturbation for each class is symmetric, $\phi_l(x,0) = \phi_l(1-x,0)$. 
The set of equations given by~\eqref{eqn:lin_approx} for $l = 1,\cdots,N_C$ has fundamental solutions of the form $\phi_l(x,t) = A_l(t)e^{ikx}$, where $k=\frac{2n\pi}{L}$ for integer $n$ and $L=1-2\epsilon$. Plugging $\phi_l(x,t) = A_l(t)e^{ikx}$ into~\eqref{eqn:lin_approx} we obtain the following system of equations 
\begin{align*}
\frac{dA_l(t)}{dt}&=2\sum_l C_{lj}\left[\left(\frac{2\sin(\epsilon k/2)}{k}-\epsilon\right)A_l(t)+\left(\frac{2\sin(\epsilon k/2)}{k} -\frac{\sin(\epsilon  k)}{ k}\right)A_j(t)\right],
\end{align*}
for $ l = 1,\cdots, N_C$, which can be written in matrix form as 
\begin{equation}
\label{eqn:eigen_ode}
\frac{ dA}{dt} = 2\epsilon MA,
\end{equation}
where 
\begin{equation}
M_{lj} = 
\begin{cases}
\sum_i C_{li}\left(\frac{2\sin(\epsilon k/2)}{\epsilon k}-1\right) +  C_{ll}\left(2\frac{\sin(\epsilon  k/2)}{\epsilon  k} -\frac{\sin(\epsilon  k)}{\epsilon k} \right) & j=l, \\
 C_{lj}\left(\frac{2\sin(\epsilon  k/2)}{\epsilon  k} -\frac{\sin(\epsilon  k)}{\epsilon k} \right) & j\neq l.
\end{cases}
\end{equation}
Fundamental solutions of~\eqref{eqn:eigen_ode} are $A(t) = ve^{2\epsilon\lambda t}$ where $\lambda$ is an eigenvalue of $M$ and $v$ is the corresponding eigenvector.

For two classes, the matrix in~\eqref{eqn:eigen_ode} is given by
\begin{equation*}
M = M(\tilde{k}) 
=
\begin{pmatrix}
(C_{11}+C_{12})f(\tilde{k})+C_{11}g(\tilde{k})  &  C_{12}g(\tilde{k}) \\
 C_{21}g(\tilde{k}) & \left(C_{21} + C_{22}\right)f(\tilde{k}) + C_{22}g(\tilde{k})
\end{pmatrix},
\end{equation*}
where
\begin{equation*}
f(\tilde{k}) = \left(\frac{2\sin(\tilde{k}/2 )}{\tilde{k}}-1\right), 
\qquad
g(\tilde{k}) = \left(\frac{2\sin(\tilde{k}/2)}{\tilde{k}} -\frac{\sin(\tilde{k})}{\tilde{k}}\right), \qquad \tilde{k} = \epsilon k.
\end{equation*}
The eigenvalues of $M$ are
\begin{equation}
\label{eqn:lambda}
\lambda_{\pm}(\tilde{k}) = \Tr(M) \pm \sqrt{\left[\Tr(M)\right]^2-4\det(M)},
\end{equation}
where $\Tr(M) = \gamma f(\tilde{k}) + \zeta g(\tilde{k})$ and $\det(M) = \alpha f(\tilde{k})^2 + \beta f(\tilde{k}) g(\tilde{k}) + (\beta - \alpha) g(\tilde{k})^2$ for non-negative constants 
\begin{align*}
\gamma &= C_{11}+C_{12}+C_{21}+C_{22},\\ 
\zeta &= C_{11}+C_{22}, \\
\alpha &= (C_{11}+C_{12})(C_{21}+C_{22}), \\
\beta &= C_{11}(C_{21} + C_{22})+C_{22}(C_{11}+C_{12}).
\end{align*}
For the degree-based {mean-field} equations all coefficients $C_{lj}$ are positive, in which case $\gamma, \zeta, \alpha$ and $\beta$ are strictly positive.
The eigenvalues are real and distinct when the term under the square-root in { Eq.~\eqref{eqn:lambda}} is positive, which is always true for the two-{class} problem (see Appendix~\ref{sec:app:realanddistinct}).  Furthermore, both eigenvalues are negative for $\tilde{k}>5$ (see Appendix~\ref{sec:app:negative}), so when seeking the maximum growth rate we only need to consider $\tilde{k} \in [0,5]$.

\subsection{Two-class example}\label{sec:twoclasslinstab}
\begin{figure}[ht]
\centering
\includegraphics[width=7.0cm]{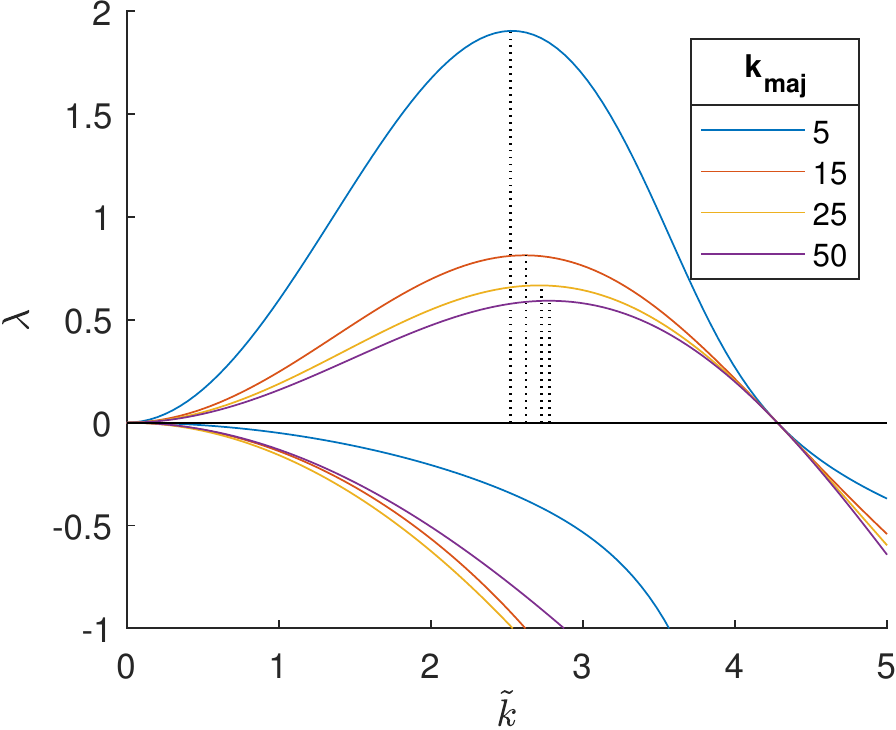}
\caption{\label{fig:lambda}Eigenvalues of $M(\tilde{k})$ for the degree-based {mean-field} equation with $10\%$ {degree-100} nodes and $90\%$ of nodes with degree $k_{maj}$. The dotted lines indicate the location where $\lambda(\tilde{k})$ is at a maximum, corresponding to the fastest growing mode.}
\end{figure}

\begin{figure}[ht]
\centering
\begin{tikzpicture}
\node at (0,0) {\includegraphics[width=8cm]{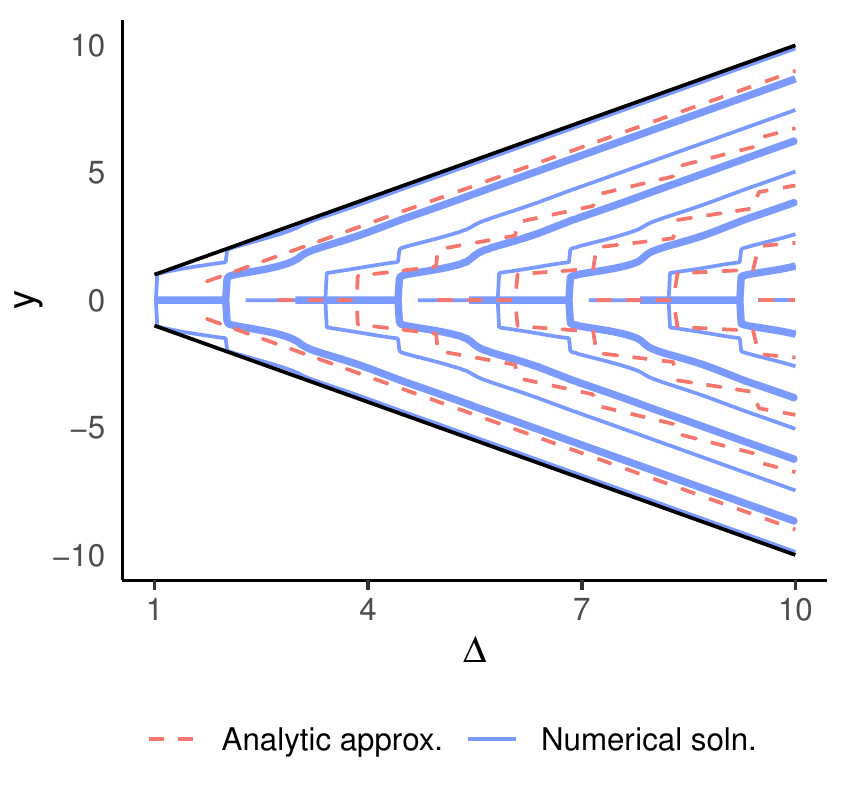}};
\node at (8,0.5) {\includegraphics[width=6.5cm]{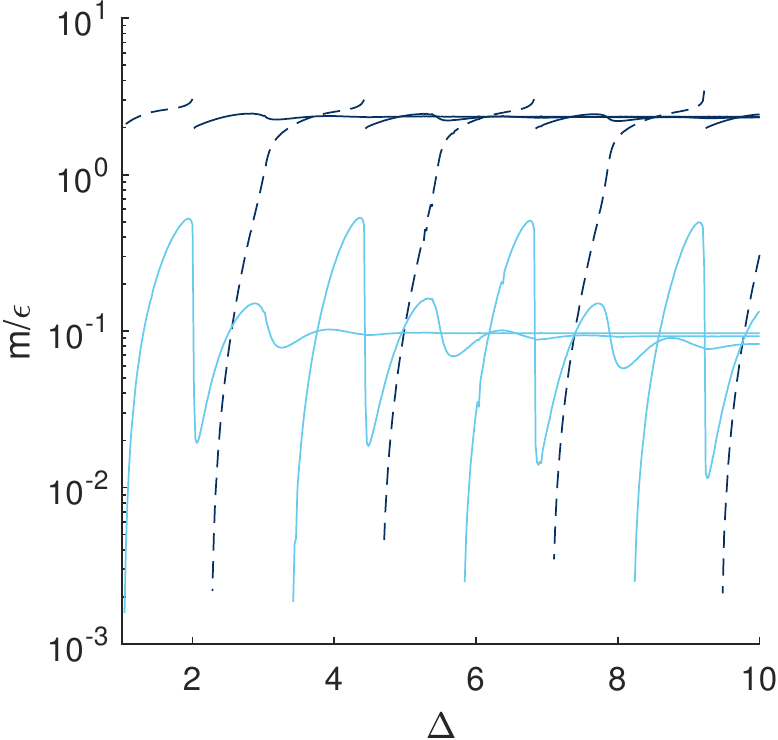}};
\node at (-4,3) {(a)};
\node at (4.5,3) {(b)};
\end{tikzpicture}
\caption{\label{fig:bif_diag_dbmf} (a) Final cluster locations predicted by the linear stability analysis (red) and numerics (blue) for the degree-based mean-field equations with $90\%$ of nodes with {degree 5} and $10\%$ of nodes with degree 100. For visualization purposes we have made the change of variable $\Delta = 1/2\epsilon,\, y = (2x-1)/2\epsilon$. The thick lines are clusters which contain both degree-100 and {degree-5} nodes and have a large mass, while the thin lines are clusters which contain only {degree-5} nodes and whose mass is an order of magnitude smaller. (b) Mass of the clusters predicted by~\eqref{eq:chapter2} when $90\%$ of nodes have degree 5 and $10\%$ of nodes have degree 100. The dark blue lines are the clusters containing both {degree-5} and {degree-100} nodes, while the light blue lines are the clusters containing only {degree-5} nodes. The mass of each cluster is proportional to $\epsilon$ and therefore the clusters get very small as $\epsilon$ gets small ($\Delta$ gets large). For this reason we plot the mass divided by $\epsilon$. The mass of the central cluster, which appears and disappears as $\epsilon$ changes, is plotted as a dashed line.}
\end{figure}

We now focus on the example of a network with two degree classes, a minority ($10\%$) of which have degree $100$ and a majority (90\%) with degree $k_{maj}$. Figure~\ref{fig:lambda} shows $\lambda(\tilde{k})$ for $k_{maj} = 5,15,25$ and 50. The location of the peak only changes slightly when the degree of the majority changes.  The cluster locations predicted by the linear stability analysis for $k_{maj} = 5$ are shown in Fig.~\ref{fig:bif_diag_dbmf}a together with the final cluster locations in the numerical solutions. We have plotted two types of clusters from the numerical solution: the thick lines are the clusters which contain {degree-5} nodes and {degree-100} nodes; the thin lines correspond to clusters which contain only {degree-5} nodes. The clusters with only {degree-5} nodes are much smaller than the clusters containing both types of nodes; the mass of the larger clusters are between 4 and 20 times the mass of smaller clusters, as illustrated in Fig.~\ref{fig:bif_diag_dbmf}b. 

The analytic approximation quite accurately predicts the location of the major clusters for $\Delta <6 $. For larger values of $\Delta$ it is accurate for clusters close to the boundaries of the opinion space, but becomes less accurate close to $y=0$. The approximation does not pick up the minor clusters, however these are an order of magnitude smaller than the major clusters and we do not expect to capture these effects from a first-order approximation.  
The ability of a linear stability analysis to correctly predict the late-time behaviour in this system may be {further} limited by the occurrence of different behaviours on different timelines, as outlined in \S\ref{sec:cbmf_numeric_approx} and \S\ref{sec:newsub2}.

\section{Discussion} \label{sec:mf_analysis_conclusions}
In this paper we have used a range of mathematical techniques to determine the early-time behaviour and the final cluster locations for the Deffuant model with confidence bound $\epsilon$. The analysis was applied to both the single mean-field equation derived by Ben-Naim \etal~\cite{Ben-Naim2003} for a well-mixed population (fully connected network) and to the corresponding equations for degree-based classes derived in~\cite{Fennell2021}. In these models, an initially symmetric opinion distribution remains symmetric, and if the opinion distribution is initially uniform then changes in opinion are driven by the presence of boundaries. In particular, at any point $x^* \in [0,\epsilon)$, there is an imbalance between interactions with people of opinions $x < x^*$ and people of opinions $x>x^*$, which leads to an overall migration in opinion to larger $x$.  An analogous argument can be made at $x^* \in (1-\epsilon,1]$. As $\epsilon$ decreases, the range of opinion space over which opinion differences are felt  is more limited, altering the opinion trajectory and leading to the formation of opinion clusters.

In the first half of the paper we focused on the limit where the confidence bound $\epsilon$ is small.  We demonstrated that the limited interaction range in the governing equations~\eqref{eq:chapter2} and~\eqref{eqn:omf} has two major consequences for opinion evolution. First, opinions evolve slowly, over a timescale $t \sim 1/\epsilon$.  Secondly, opinion evolution initially occurs only in a region of size $\epsilon$ near each boundary, and the density of opinions does not change in the central region of opinion space until late time.  

Numerical solution of the governing equations illustrates that the {earliest changes in the opinion distribution occur} near the boundaries; this is associated with the limited range of interaction there.  Far from the boundaries the opinion density remains fixed at $P \approx 1$ until late times.  These observations motivate a rescaling of the mean-field equations to obtain an approximate model which is independent of $\epsilon$.  The approximate model must be solved numerically, but the solution can then be rescaled to approximate the distribution for any value of $\epsilon$. 

In the single-class case, the small-$\epsilon$ model predicts that all mass moves out of the boundary regions into the central region, {as observed in} numerical simulations of the Deffuant model on fully connected networks \cite{Deffuant2000}. For the degree-based classes, the analysis was carried out for an arbitrary number of classes, and the approximate model was then solved numerically for a network with two degree classes, which was one of the examples used in~\cite{Fennell2021}.  Deviations between the full and asymptotic solutions for a system of two classes with degrees 5 and 100 point to the emergence of two distinct evolution timescales in the limit where classes have very different degrees; this leads to an overall increase in the movement of mass towards the centre of opinion space.  Our approximation for the network with two degree classes also demonstrates that the mass {remaining} in the boundary regions depends on the degree $k_{maj}$, which is in agreement with  results from the numerical solution of the full class-based mean-field equations as described in~\cite{Fennell2021}.

In the second part of the paper we carried out a linear stability analysis of the mean-field equations in the central region to predict both the number and location of opinion clusters in the $t\to \infty$ limit. The linear stability analysis gave an accurate prediction of the location of major clusters, i.e., those with mass of the order of $\epsilon$. It did not, however, capture the occurrence of so-called minor clusters, which are an order of magnitude smaller than the major clusters.
An important assumption of the linear stability analysis is that the initial distribution, and hence the distribution at all times, is symmetric.  Our prediction for the number of clusters would not change if this assumption was removed, but it is required to predict the location of the clusters.

We conclude by commenting on some open questions and areas for further exploration. Firstly, the analysis in this paper addressed the specific situation where opinions were updated to the average of an interacting pair of agents $i$ and $j$.  This can be extended to a generalized averaging rule $x_{i/j} = (1-\mu)x_{i/j} + \mu x_{j/i}$, and the corresponding analysis{---which is qualitatively unchanged from the $\mu=1/2$ case studied here---}is reported in~\cite{FennellThesis}.
The small-$\epsilon$ analysis could be repeated for a non-uniform initial distribution of the form $P(x,0) = f(x/\epsilon)$, as the change of variables $x = \epsilon\tilde{x}$ will result in a distribution $\tilde{P}(\tilde{x},0) = f(\tilde{x})$ which is independent of $\epsilon$. In the case of the linear stability analysis, however, we require a uniform initial distribution in the central region, i.e.{,} a steady state.  The boundary regions are not included in the analysis, and small changes in the initial conditions there would not affect our results. 
Finally, in this paper we presented predictions for a case study with two degree classes, but analogous predictions could be made for other class-based systems.  As an example, dividing a population into different classes with different update rules could be used to model ``strong leaders'' or other subgroups who exert strong influence on those around them~\cite{During2009}.

\section*{Acknowledgements}
This work is partly supported by Science Foundation Ireland under grant numbers 16/IA/4470 (A.D. and J.G), 12/RC/2289 P2 (J.G.) and 16/RC/3918 (J.G.), and by the Irish Research Council (S.F.). 

\bibliographystyle{siamplain}
\bibliography{M149976}

\appendix
\section{Growth rates for two-class model}
\subsection{Real and distinct eigenvalues}\label{sec:app:realanddistinct}
We need to prove
\begin{equation}
\label{eqn:distinct_eigen}
\left[\Tr(M)\right]^2-4\det(M)=\left(\tilde{\gamma}f(\tilde{k})+\tilde{\zeta}g(\tilde{k})\right)^2 + 4C_{12}C_{21}g(\tilde{k})^2>0,
\end{equation}
where $\tilde{\zeta} = C_{11}-C_{22}$ and $\tilde{\gamma} = C_{11}+C_{12}-C_{21}-C_{22}$. Equation~\ref{eqn:distinct_eigen} holds when $\tilde{\gamma}f(\tilde{k})+\tilde{\zeta}g(\tilde{k})$ is non-zero or $C_{12}, C_{21}$ and $g(\tilde{k})$ are all non-zero. 
We will assume $\tilde{\gamma}\neq 0$ since for the degree-based {mean-field (MF)} equations $\tilde{\gamma}=0$ only when the degree of nodes in both classes are the same, in which case we can use the original MF equation. Since $C_{12}>0$ and $C_{21}>0$ we have $C_{12}C_{21}g(\tilde{k})= 0$ only  when $g(\tilde{k})=0$. In this case $\tilde{\gamma}f(\tilde{k})+\tilde{\zeta}g(\tilde{k}) = \tilde{\gamma}f(\tilde{k})$ which is non-zero for $\tilde{k}>0$ since $f(\tilde{k})<0$. Thus for the degree-based MF equations with two classes the eigenvalues are real and distinct.

\subsection{Limited range for positive eigenvalues}\label{sec:app:negative}
For $\tilde{k}>5$ we have $f(\tilde{k}) <0$ and $f(\tilde{k})+g(\tilde{k})<0$ and so
\begin{align*}
\Tr(M) &= \gamma f(\tilde{k}) + \zeta g(\tilde{k})\\
&=(\gamma-\zeta) f(\tilde{k}) + \zeta\left( f(\tilde{k})+g(\tilde{k})\right)\\
&< 0,
\end{align*}
and 
\begin{align*}
\det(M) &=\alpha f(\tilde{k})^2 + \beta f(\tilde{k}) g(\tilde{k}) + (\beta - \alpha) g(\tilde{k})^2 \\
&= \left(f(\tilde{k})+g(\tilde{k})\right)\left[(2\alpha-\beta)f(\tilde{k})+(\beta -\alpha)\left(f(\tilde{k})+g(\tilde{k})\right)\right]\\
&>0,
\end{align*}
where $2\alpha  - \beta =  C_{11}C_{22} + 2C_{12}C_{21} + C_{12}C_{22} >0$ and we assume $\beta -\alpha  =  C_{11}C_{22} - C_{12}C_{21}\geq0$. (For the degree-based problem $\beta - \alpha = 0$).
Thus for $\tilde{k}>5$ the two eigenvalues are negative.
This means, as in the one dimensional case, we only need to look at $\lambda_{\pm}$ for $\tilde{k}\in[0,5]$ to find the global max. 

\end{document}